\documentclass[pdflatex,iicol]{sn-jnl}

\usepackage{graphicx}%
\usepackage{multirow}%
\usepackage{amsmath,amssymb,amsfonts}%
\usepackage{amsthm}%
\usepackage{mathrsfs}%
\usepackage[title]{appendix}%
\usepackage{xcolor}%
\usepackage{textcomp}%
\usepackage{manyfoot}%
\usepackage{booktabs}%
\usepackage{algorithm}%
\usepackage{algorithmicx}%
\usepackage{algpseudocode}%
\usepackage{listings}%
\usepackage{tikz}
\usetikzlibrary{calc}

\theoremstyle{thmstyleone}%
\theoremstyle{thmstyletwo}%

\theoremstyle{thmstylethree}%

\begin{document}

\title[A polygonal Reissner-Mindlin plate element based on the scaled boundary finite element method]{A polygonal Reissner-Mindlin plate element based on the scaled boundary finite element method}


\author*[1]{\fnm{Anna} \sur{Hellers}}\email{hellers@lbb.rwth-aachen.de}

\author[1]{\fnm{Mathias} \sur{Reichle}}\email{reichle@lbb.rwth-aachen.de}

\author[1]{\fnm{Sven} \sur{Klinkel}}\email{klinkel@lbb.rwth-aachen.de}

\affil[1]{\orgdiv{Chair of Structural Analysis and Dynamics}, \orgname{RWTH Aachen University}, \orgaddress{\street{Mies-van-der-Rohe-Str. 1}, \city{Aachen}, \postcode{52074}, \country{Germany}}}


\abstract{In this work, a polygonal Reissner–Mindlin plate element is presented. The formulation is based on a scaled boundary finite element method, where in contrast to the original semi-analytical approach, linear shape functions are introduced for the parametrization of the scaling and the radial direction. This yields a fully discretized formulation, which enables the use of non-star-convex-polygonal elements with an arbitrary number of edges, simplifying the meshing process. To address the common effect of transverse shear locking for low-order Reissner-Mindlin elements in the thin-plate limit, an assumed natural strain approach for application on the polygonal scaled boundary finite elements is derived. Further, a two-field variational formulation is introduced to incorporate three-dimensional material laws. Here the plane stress assumptions are enforced on the weak formulation, facilitating the use of material models defined in three-dimensional continuum while considering the effect of Poisson’s thickness locking. The effectiveness of the proposed formulation is demonstrated in various numerical examples.}

\keywords{Scaled Boundary Finite Element Method, Polygonal element, Reissner--Mindlin, Shear locking, Assumed Natural Strains, Voronoi tessellation}



\maketitle

\section{Introduction}\label{sec1}
Thin plate structures are of wide use in different domains of engineering. Their structural analysis is commonly performed using finite element methods (FEM). In this context, common plate elements are constrained to simple triangular and quadrilateral shapes. This restriction complicates the meshing process, especially in case of complex geometries and local mesh refinements. User intervention is often required, which compromises the computational efficiency \cite{Song2018}. Here, polygonal element formulations have proven to be beneficial. They facilitate the automatic handling of hanging nodes, local mesh refinements \cite{Zhang2020} and the meshing of a domain with voids or cracks \cite{Khoei2015}. 

The first polygonal shaped elements with linear interpolation functions have been developed after the introduction of the rational Wachspress shape functions \cite{Wachspress1971} in 1971. Nowadays, they belong to the polygonal elements based on barycentric coordinates, among which not only rational, but also logarithmic and trigonometric shape functions are used \cite{Perumal2018}. Due to their complexity, a higher order quadrature rule is often necessary \cite{Nguyen2017}. Ngyuen--Xuan presented in \cite{Nguyen2017} polygonal plate elements based on barycentric coordinates.

A different approach is used in the Virtual Element Method (VEM), first introduced in 2012 \cite{Beirao2012}. VEM based thin plate formulations can be found in ~\cite{Brezzi2013,Chinosi2018,DAltri2022}. In this method, the shape functions are not explicitly defined. Thus, the displacement field within the element domain is unknown and therefore determined through projection onto a polynomial subspace. This induces the need for stabilization methods to avoid zero energy modes \cite{Cangiani2015}. 

With the Scaled Boundary Finite Element Method (SBFEM) proposed by Song and Wolf in 1997 \cite{Song1997}, polygonal elements restricted to star shaped geometries are derived. They stand out by a scale separation of a radial and circumferential direction. Originally, the weak form of equilibrium is only applied along the element boundary, whereas in scaling direction, the strong form is enforced, yielding a semi-analytical formulation. However, the analytical approach in scaling direction poses a significant challenge to solving of nonlinear problems. In \cite{Klinkel2019}, a fully discretized SBFEM membrane formulation is presented, which allows an independent definition of shape functions in radial and in circumferential direction and the use of non-star-convex element shapes. An extension to nearly incompressible materials is provided in \cite{Sauren2023}. To derive thin plate formulations in SBFEM, Kirchoff-Love plate theory ~\cite{Kirchhoff1850} was used at first \cite{Hughes1988}. Formulations relying on this theory require a C$^{1}$-continuity across element boundaries for their shape functions, which complicates the derivations of finite elements \cite{Bischoff2018}. Here, Reissner--Mindlin theory \cite{Reissner1945,Mindlin1951} can be beneficial due to a lower continuity requirement. In difference to Kirchhoff-Love, this theory incorporates transverse shear strains.

However, it is known that especially low-order Reissner--Mindlin plate formulations suffer from transverse shear locking if the same shape functions are used for the interpolation of the displacements and the rotations. This results in an overestimation of the stresses, while the deformations are underestimated, which compromises the effectiveness of finite element formulations. Therefore, various remedies have been derived over the years, among which the selective reduced integration method is a first attempt \cite{Hughes1977,Hughes1978}. Here, transverse shear strains, being the parasitic components, are integrated using a deficient amount of Gauss-points. This leads to a rather simple approach, which successfully removes locking, but reduces the element stability due to zero energy modes \cite{Spilker1982}. 

Batoz et al. followed a different approach which enforces the zero-shear condition at discrete points \cite{Batoz1980}. Starting point is a plate formulation which considers transverse shear deformations. Though, the energy related to shear is neglected and the Kirchhoff hypothesis is introduced along the element edges, hence the name discrete Kirchhoff triangle or quadrilateral (DKT, DKQ), depending on the element shape.

In 1990, Simo and Rifai proposed an alleviation of locking through enhanced assumed strains (EAS) \cite{Simo1990}, where the  shear components that cause locking are balanced through enhancement. But, the numerical cost is high compared to other locking remedies, such as the discrete shear gap approach (DSG)\cite{Bletzinger2000}. In this framework, the gaps, being the difference between the total and bending deformations, are evaluated at the nodes, where they contain no parasitic terms. Afterwards, they are interpolated over the element domain.

Similarities to the assumed natural strain approach (ANS) can be found, see \cite{Bletzinger2000}. Originally introduced by Bathe and Dvorkin, for a bilinear quadrilateral finite plate element \cite{Bathe1985}, this approach has been widely used for the alleviation of shear locking in plates and shells, such e.g. ~\cite{Lee2010, Ko2016, Choi2024}. The shear strains are evaluated at the so-called tying points, while the curvature strains are evaluated at the nodes, resulting in a distinctive interpolation  of both components. This yields the terminology of Mixed Interpolation of Tensorial Components (MITC). 

Most of the locking remedies are limited to an application on triangular or quadrilateral element shapes and a shortage of robust remedies for polygonal elements still remains. So far, a polygonal Reissner--Mindlin element formulation has been presented by Nguyen-Xuan and Videla using barycentric coordinates ~\cite{Nguyen2017, Videla2019}. Recently, the discrete shear projection method was extended towards an application on arbitrary polygons using polygonal serendipity shape functions \cite{Akhila2025}. The transverse shear locking effects are reduced by enforcing Timoshenko's beam assumptions along the edges of the polygons. Considering the framework of VEM, transverse shear locking can be alleviated through replacement of the rotations in the variational formulation by the transverse shear strains ~\cite{Beirao2017}. 
In semi-analytical SBFEM, a polygonal element has been derived using the discrete Kirchhoff theory, see ~\cite{Dieringer2011,Li2021}.  Meanwhile, there remains a lack of thin plate formulations based on Reissner--Mindlin plate theory in the framework of fully-discretized SBFEM.  

Considering the material laws of plate formulations, they are usually limited to a two dimensional description by enforcement of the plane stress assumptions on the strong form. However, an incorporation of three dimensional laws is of interest, especially in case of more complex material models, since they are usually expressed in $3D$-continuum \cite{Klinkel2002}. For the incorporation, an enhancement of the thickness strains based on the enhanced assumed strain method (EAS) can be used ~\cite{Betsch1996,Klinkel2008}. Its variational formulation relies on the Green-Lagrange strain tensor and the second Piola Kirchhoff tensor. To avoid Poisson`s thickness locking in case of non-zero Poisson`s ratios a linear interpolation of the strains in thickness direction is required \cite{Willmann2023}.

In this contribution, a polygonal Reissner--Mindlin plate formulation is presented using fully-discretized SBFEM. The shape functions are chosen linear in scaling and radial direction, resulting in a low order element formulation. To address the shear locking effects, an ANS approach is derived for arbitrarily polygonal element shapes. The present approach is applicable on any so-called SBFEM section. Further, both two and three-dimensional material laws are considered by enforcing the plane stress assumptions first on the strong form and then on the weak form. 

The presented work consists of 9 sections. In Section \ref{sec: Governing equations}, the governing equations of the plate bending problem are introduced. It follows the derivation of the weak form of equilibrium in Section \ref{sec: Weak form of equilibrium}. To solve the variational formulation by numerical methods, the SBFEM parameterization is defined in Section \ref{sec: Scaled boundary finite element parameterization}. This allows the introduction of the corresponding numerical approximation in Section \ref{sec: Numerical approximation}. In Section \ref{sec: Assumed natural strain field}, the ANS approach is derived to overcome transverse shear locking. Then, the $3D$-material laws are incorporated in Section \ref{sec: Incorporation of 3D material laws}. Section \ref{sec: Numerical examples} examines the formulation with various benchmarking examples and a brief conclusion and outlook are given in Section \ref{sec: Conclusion and outlook}.

\section{Governing equations} \label{sec: Governing equations}
Starting point for the derivation of the element formulation is the definition of the basic equations, which consist of kinematics, material laws and equilibrium. 

\begin{figure}
   \centering
   \input{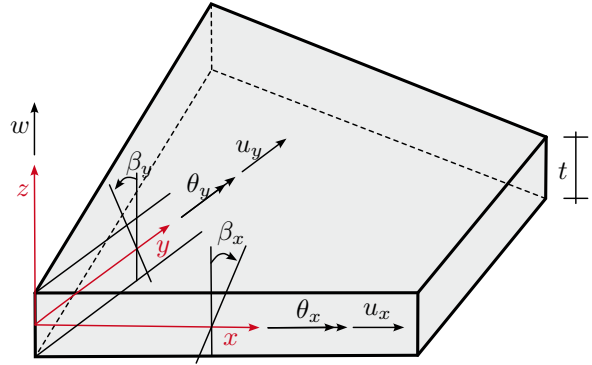}
    \caption{Visualization of the degrees of freedom of the plate formulation with $3D$-material laws.}
   \label{fig: 1}
\end{figure}

First, the kinematics are introduced. The plate formulation incorporates three degrees of freedom being the vertical displacement $w$ in $z$-direction and the rotations $\beta_{x}$ and $\beta_{y}$, around the $x$- and the $y$-direction, respectively. Consistent with the plane stress assumption, the strains in thickness direction are neglected, implying that the out of plane displacement $w(x,y)$ is independent of the thickness coordinate. Applying the assumption directly on the constitutive laws, reduces the material laws to the two-dimensional space. Meanwhile, a consideration of full $3D$-material laws can be beneficial and is therefore introduced in Section \ref{sec: Incorporation of 3D material laws}. For this, the membrane degrees of freedom $u_{x}$, $u_{y}$ are necessary, yielding a total of five degrees of freedom per node, as depicted in Fig.\ref{fig: 1}.   

In the framework of Reissner--Mindlin plate theory, the curvature $\boldsymbol{\kappa}$ and transverse shear strains $\boldsymbol{\gamma}$ have to be considered. They both depend on the rotations $\beta_{x}$ and $\beta_{y}$, and the vertical displacements $w$ and are defined as 
\begin{equation}
    \boldsymbol{\kappa} = \begin{bmatrix}
        \beta_{x,x}\\
        -\beta_{y,y}\\
        \beta_{x,y} - \beta_{y,x}    
    \end{bmatrix} = \begin{bmatrix}
        0 & \frac{\partial}{\partial x} & 0 \\
        0 & 0 & -\frac{\partial}{\partial y}\\
        0 & \frac{\partial}{\partial y} & -\frac{\partial}{\partial x}
    \end{bmatrix}
   \begin{bmatrix}
        w\\
        \beta_{x}\\
        \beta_{y}
    \end{bmatrix} = \mathbf{D}_{b}\mathbf{v}
    \label{Eq: Curvature Strains}
\end{equation}
and 
\begin{equation}
    \boldsymbol{\gamma} = \begin{bmatrix}
        w{,x} + \beta_{x}\\
        w{,y} - \beta_{y}
    \end{bmatrix} = \begin{bmatrix}
        \frac{\partial}{\partial x} & 1& 0\\
        \frac{\partial}{\partial y} & 0 & -1
    \end{bmatrix}
    \begin{bmatrix}
        w\\
        \beta_{x}\\
        \beta_{y}
    \end{bmatrix} = \mathbf{D}_{s}\mathbf{v}.
    \label{Eq: Shear strain}
\end{equation}
If three dimensional material laws are incorporated, the membrane strains $\boldsymbol{\varepsilon}$ are additionally required. They are written as
\begin{equation}
    \boldsymbol{\varepsilon} = \begin{Bmatrix}
        u_{x,x} \\ u_{y,y} \\ u_{x,y} + u_{y,x}
    \end{Bmatrix}=\begin{bmatrix}
        \frac{\partial}{\partial x} & 0 \\ 0 & \frac{\partial}{\partial y} \\ \frac{\partial}{\partial y} & \frac{\partial}{\partial x}
    \end{bmatrix}\begin{Bmatrix}
        u_{x}\\ u_{y}
    \end{Bmatrix} = \mathbf{D}_{m}\mathbf{u}.
    \label{Eq: Membrane Strains}
\end{equation} 
The strains can be summarized in the geometric strain vector $\boldsymbol{\epsilon}_{g}$, which reads
\begin{equation}
     \boldsymbol{\epsilon}_{g} = \begin{bmatrix}
        \mathbf{D}_{m} & \mathbf{0}\\ \mathbf{0} & \mathbf{D}_{b} \\ \mathbf{0} & \mathbf{D}_{s}
    \end{bmatrix}\begin{bmatrix}
        \mathbf{u} \\ \mathbf{v}
    \end{bmatrix} = \mathbf{D}\mathbf{d}\label{eq: geometrical strains},
\end{equation}
including membrane strains.
It depends on the differential operators $\mathbf{D}$ and the deformation vector  $\mathbf{d} = \begin{bmatrix}
    \mathbf{u} & \mathbf{v}
\end{bmatrix}^\text{T}$. 

Next, the material laws are introduced. In case of linear elastic material behavior, the membrane, bending and shear stresses, $\mathbf{n}$, $\mathbf{m}$ and $\mathbf{q}$ are expressed as

\begin{equation}
    \mathbf{n} = \mathbf{C}_{m}\boldsymbol{\varepsilon} \qquad \mathbf{m} = \mathbf{C}_{b}\boldsymbol{\kappa} \qquad \mathbf{q} = \mathbf{C}_{s} \boldsymbol{\gamma}, \label{eq: Stresses}
\end{equation}
with 
\begin{align}
        &\mathbf{C}_{m} = \frac{Et}{1-\nu^2}\begin{bmatrix}
        1&\nu&0\\ \nu&1&0 \\ 0&0&\frac{1-\nu}{2}
    \end{bmatrix}, \quad \mathbf{C}_{b} = \frac{t^2}{12}\mathbf{C}_{m} , \\ &\mathbf{C}_{s} = \frac{Etk}{2(1+\nu)}\begin{bmatrix}
        1&0\\0&1
    \end{bmatrix},
\end{align}
where $E$ is the Young's modulus, $\nu$ is the Poisson's ratio, $k$ is the shear correction factor set to $5/6$ and $t$ is the thickness of the plate.

The global stresses are defined as $\hat{\boldsymbol{\sigma}} = \mathbf{C}\boldsymbol{\epsilon}_{g}$, with 
\begin{equation}
     \mathbf{C} = \begin{bmatrix}
        \mathbf{C}_{m} & \mathbf{0} &\mathbf{0}\\
        \mathbf{0} & \mathbf{C}_{b} & \mathbf{0} \\
        \mathbf{0} & \mathbf{0} & \mathbf{C}_{s}
    \end{bmatrix}.
\end{equation}
Last, the static equilibrium has to be fulfilled.

To apply numerical methods, the weak form of equilibrium has to be derived. First, the weak form is given for a plate formulation with $2D$-material laws. To extent the applicability, $3D$-material laws under the assumption $\sigma_{z} = 0$ are incorporated in Section \ref{sec: Incorporation of 3D material laws}. 
\section{Weak form of equilibrium} \label{sec: Weak form of equilibrium}
In case of $2D$-material laws, a one field variational formulation is solved. Here, the kinematics and material laws are exactly fulfilled, whereas the equilibrium equation holds in an average sense. The weak form of equilibrium is derived through multiplication of the strong form by the test function $\delta \mathbf{d}$. Then integration by parts is applied over the midsurface $\mathit{\Omega}$ of the plate, yielding 
\begin{multline}
    \underbrace{\int_{\mathit{\Omega}}(\delta\boldsymbol{\epsilon}_{g})^{\text{T}}\hat{\boldsymbol{\sigma}}\,\text{d}A}_{(*)} -\int_{\mathit{\Omega}}\delta\mathbf{d}^{\text{T}} \mathbf{b}\,\text{d}A \\-\int_{\partial \mathit{\Omega}}\delta\mathbf{d}^{\text{T}}\bar{\mathbf{t}}\,\text{d}S = 0 \quad \text{in} \quad \mathit{\Omega},\label{eq: Weak form}
\end{multline}
where the integral $(*)$ can be rewritten as 
\begin{multline}
\delta\mathbf{u}^{\text{T}}\int_{\mathit{\Omega}}\mathbf{D}_{m}^{\text{T}}\mathbf{C}_{m}\mathbf{D}_{m}\,\text{d}A \mathbf{u} +
\delta\mathbf{v}^{\text{T}}\int_{\mathit{\Omega}}\mathbf{D}_{b}^{\text{T}}\mathbf{C}_{b}\mathbf{D}_{b}\,\text{d}A \mathbf{v} \\+\delta\mathbf{v}^{\text{T}}\int_{\mathit{\Omega}}\mathbf{D}_{s}^{\text{T}}\mathbf{C}_{s}\mathbf{D}_{s}\,\text{d}A \mathbf{v}
\end{multline}
and $\mathbf{b}$ is a surface load and $\bar{\mathbf{t}}$ is a traction force.
\section{Scaled boundary finite element parameterization} \label{sec: Scaled boundary finite element parameterization}

\subsection{Discretization of a polygonal domain}
\begin{figure*}
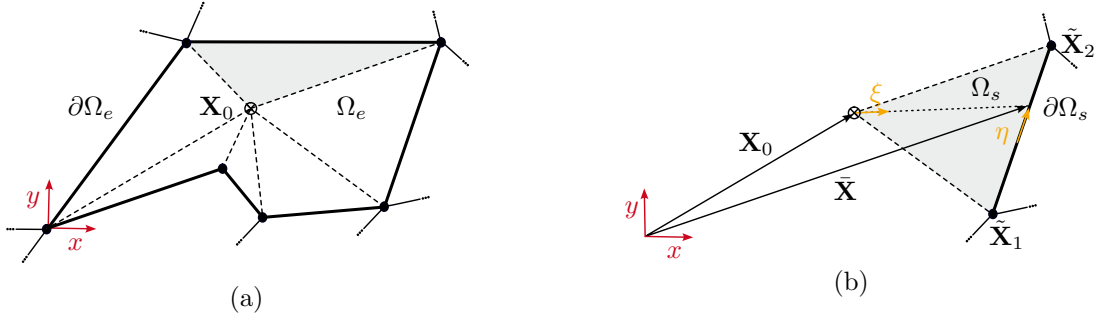

\centering
\begin{minipage}{0.49\linewidth}
\input{Fig_2_a.tex}
    \centering (a)
\end{minipage}
\begin{minipage}{0.49\linewidth}
    \input{Fig_2_b.tex}
    \centering (b)
\end{minipage}
\caption{Illustration of a polygonal element domain $\Omega_{e}$ (a) consisting of six sections, the scaling center with its position vector $\mathbf{X}_{0}$ and the element boundary $\partial\Omega_{e}$ and illustration of a section $\Omega_{s}$ (b) consisting of the scaling center with its position vector $\mathbf{X}_{0}$, the section boundary $\partial\Omega_{s}$, the boundary nodes $\tilde{\mathbf{X}}_{i}$, the position vector $\bar{\mathbf{X}}$ of a node at the boundary and the parameterization ($\xi,\eta$).}
\label{fig: 2}
\end{figure*}
In SBFEM, the domain $\Omega$ is discretized using $n_{e}$ polygonal elements $\Omega_{e}$, such that $\Omega = \cup^{n_{e}}_{e=1}\Omega_{e}$. Each element possesses an element boundary $\partial\Omega_{e}$ and a scaling center, from which every point of the element is visible and therefore the scaling requirement is fulfilled. In accordance to the concept of traditional SBFEM \cite{Song2018}, the boundary $\partial\Omega_{e}$ is scaled to the scaling center to cover the whole element domain $\Omega_{e}$. This domain is defined as assemblage of triangular sections $\Omega_{s}$, such that $\Omega_{e} = \cup^{n_{s}}_{e=1}\Omega_{s}$. In analogy, the boundary of the element $\partial \Omega_{e}$ consists of $n_{s}$ line elements being the boundaries $\partial \Omega_{s}$ of the sections. These line elements are described by the local parameter $\eta \in [-1,1]$ in circumferential direction and scaled to the scaling center with the scaling parameter $\xi \in [0,1]$ in radial direction. As in \cite{Klinkel2019}, independent interpolations are introduced in both directions, yielding a fully discretized SBFEM formulation that differs from the semi-analytical approach of original SBFEM formulations ~\cite{Song2018, Song1997}. An example of a scaled boundary element can be seen in Fig.\ref{fig: 2}(a), while a section and its parameterization is displayed in Fig.\ref{fig: 2}(b).

Using SBFEM, the position vector $\bar{\mathbf{X}}$ of a point located at a section's boundary $\partial \Omega_{s}$ is calculated from 
\begin{equation}
\bar{\mathbf{X}}(\eta) = \bar{\mathbf{N}}(\eta) \tilde{\mathbf{X}} \quad \text{on} \quad \partial\Omega_{s}, \label{eq: Position vector boundary node}
\end{equation}
where $\tilde{\mathbf{X}}$ is a vector containing the nodal coordinates $\tilde{\mathbf{X}}_{1}$ and $\tilde{\mathbf{X}}_{2}$. $\bar{\mathbf{N}}$ is the matrix containing the linear shape functions $\bar{N}_{1} = \frac{1}{2}(1-\eta)$ and $\bar{N}_{2} = \frac{1}{2}(1+\eta)$ at the boundary, such that
\begin{equation}
   \bar{\mathbf{N}}(\eta) =\begin{bmatrix}
        \bar{N}_{1} & 0 &  \bar{N}_{2}& 0\\ 0 &\bar{N}_{1} & 0 &  \bar{N}_{2}
    \end{bmatrix}.
\end{equation}
The coordinates $\bar{\mathbf{X}}$ of the point at a boundary are then used to derive the coordinates $\mathbf{X}$ of a node within a section $\Omega_{s}$ with
\begin{equation}
\mathbf{X}(\xi,\eta) = \mathbf{X}_{0} + \xi (\bar{\mathbf{X}}(\eta) - \mathbf{X}_{0}) \quad \text{in} \quad \Omega_{S}.
\end{equation}
Here, $\mathbf{X}_{0}$ are the coordinates of the scaling center.

The mapping between the physical and the parameter space is proceeded using the Jacobian matrix being defined as
\begin{equation}
    \mathbf{J} = \begin{bmatrix}
        1 & 0 \\ 0 & \xi
    \end{bmatrix} \bar{\mathbf{J}},
\end{equation}
with $\bar{\mathbf{J}}$ being the Jacobian matrix on the boundary $\partial \Omega_{s}$, given by
\begin{equation}
    \bar{\mathbf{J}} = \begin{bmatrix}
        (\bar{\mathbf{X}} - \mathbf{X}_{0})^{\text{T}} \\ \bar{\mathbf{X}}_{,\eta}^{\text{T}} 
    \end{bmatrix} = \begin{bmatrix}
        \bar{\mathbf{G}}_{1}^{\text{T}}(\eta) \\ \bar{\mathbf{G}}_{2}^{\text{T}}(\eta)
    \end{bmatrix},
\end{equation}
where $\bar{\mathbf{G}}_{\alpha}(\eta)$, with $\alpha, \beta \in \{1,2\}$, are the covariant basis vectors. Next, the orthogonality condition $\mathbf{G}_{\alpha}\cdot \mathbf{G}^{\beta} = \delta_{\alpha}^{\beta}$, with the Kronecker-delta $\delta_{\alpha}^{\beta}$, is applied to derive the contravariant basis vectors $\bar{\mathbf{G}}^{\alpha}$, such that 
\begin{subequations}
    \begin{align}
        &\bar{\mathbf{G}}^{1}(\eta) = \frac{1}{\text{det}(\bar{\mathbf{J}})}\begin{bmatrix}
        \bar{y}(\eta)_{,\eta}\\
        -\bar{x}(\eta)_{,\eta}
        \end{bmatrix},\\
         &\bar{\mathbf{G}}^{2}(\eta) = \frac{1}{\text{det}(\bar{\mathbf{J}})}\begin{bmatrix}
            -(\bar{y}(\eta) - y_{0})\\
            (\bar{x}(\eta) - x_{0})
        \end{bmatrix}.
    \end{align}\label{eq: Contravariant basis}
\end{subequations}

Using Eq.\eqref{eq: Contravariant basis}, the inverse of the Jacobian $\mathbf{J}$ can be derived to
\begin{equation}
    \mathbf{J}^{-1} = \begin{bmatrix}
        \bar{\mathbf{G}}^{1}(\eta) & \bar{\mathbf{G}}^{2}(\eta)
    \end{bmatrix} \begin{bmatrix}
        1 & 0 \\ 0 & \frac{1}{\xi}
    \end{bmatrix}. \label{eq: inverse of jacobian}
\end{equation}
As a result, the partial derivatives with respect to the physical coordinates $x$ and $y$ in the framework of SBFEM are written as
\begin{subequations}
    \begin{align}
          &\frac{\partial}{\partial x} = \bar{\mathbf{G}}^{1}(\eta) \cdot \mathbf{e}_{1}\frac{\partial}{\partial\xi} + \frac{1}{\xi} \bar{\mathbf{G}}^{2}\cdot \mathbf{e}_{1}(\eta)\frac{\partial}{\partial\eta}, \\ &\frac{\partial}{\partial y} = \bar{\mathbf{G}}^{1}(\eta) \cdot \mathbf{e}_{2}\frac{\partial}{\partial\xi} + \frac{1}{\xi} \bar{\mathbf{G}}^{2}\cdot \mathbf{e}_{2}(\eta)\frac{\partial}{\partial\eta}. 
    \end{align}\label{eq: Partial derivatives}
\end{subequations}
\subsection{Parametrization of the kinematics}
Using the partial derivatives from Eq.\eqref{eq: Partial derivatives}, the differential operators $\mathbf{D}_{m}$,  $\mathbf{D}_{b}$ and $\mathbf{D}_{s}$ can be derived as
\begin{subequations}
    \begin{align}
         &\mathbf{D}_{m} = \mathbf{m}_{1}(\eta)\frac{\partial}{\partial \xi} + \frac{1}{\xi}\mathbf{m}_{2}(\eta)\frac{\partial}{\partial \eta}, \\ &\mathbf{D}_{b} = \mathbf{b}_{1}(\eta)\frac{\partial}{\partial \xi} + \frac{1}{\xi}\mathbf{b}_{2}(\eta)\frac{\partial}{\partial \eta}, \\ &\mathbf{D}_{s} = \mathbf{s}_{1}(\eta)\frac{\partial}{\partial \xi} + \frac{1}{\xi}\mathbf{s}_{2}(\eta)\frac{\partial}{\partial \eta} +\mathbf{s}_{3}.
    \end{align}
\end{subequations}
Here, the matrices $\mathbf{m}_{\alpha}$, $\mathbf{b}_{\alpha}$ and $\mathbf{s}_{\alpha}$, with $j=1,2$ are defined as
\begin{subequations}
    \begin{align}
         &\mathbf{m}_{j} = \begin{bmatrix}
            \bar{\mathbf{G}}^{\alpha}(\eta) \cdot \mathbf{e}_{1} & 0\\
            0 & \bar{\mathbf{G}}^{\alpha}(\eta) \cdot \mathbf{e}_{2}\\
            \bar{\mathbf{G}}^{\alpha}(\eta) \cdot \mathbf{e}_{2} & \bar{\mathbf{G}}^{\alpha}(\eta) \cdot \mathbf{e}_{1}
        \end{bmatrix}, \\ & \mathbf{b}_{j} =
        \begin{bmatrix}
            0 & \bar{\mathbf{G}}^{\alpha}(\eta) \cdot \mathbf{e}_{1} & 0\\
            0 & 0 & -\bar{\mathbf{G}}^{\alpha}(\eta) \cdot \mathbf{e}_{2}\\
            0 & \bar{\mathbf{G}}^{\alpha}(\eta) \cdot \mathbf{e}_{2} & -\bar{\mathbf{G}}^{\alpha}(\eta) \cdot \mathbf{e}_{1}
        \end{bmatrix}, \\ & \mathbf{s}_{\alpha} = \begin{bmatrix}
             \bar{\mathbf{G}}^{\alpha}(\eta) \cdot \mathbf{e}_{1} & 0 & 0\\
            \bar{\mathbf{G}}^{\alpha}(\eta) \cdot \mathbf{e}_{2}&0&0\\
        \end{bmatrix},
    \end{align}
    and depend on the local parameter $\eta$ at the boundary, while $\mathbf{s}_{3}$ is constant and written as
    \begin{equation}
        \mathbf{s}_{3}= \begin{bmatrix}
        0&1&0\\
        0&0&-1
    \end{bmatrix}.
    \end{equation} \label{eq: m,b,s}
\end{subequations}
Introducing Eq.\eqref{eq: m,b,s} into the membrane, curvature and shear strain expressions gives
\begin{subequations}
   \begin{align}
&\boldsymbol{\varepsilon} = \mathbf{m}_{1}(\eta)\frac{\partial \mathbf{u}}{\partial \xi }+\frac{1}{\xi}\mathbf{m}_{2}(\eta)\frac{\partial \mathbf{u}}{\partial \eta } \label{eq: membrane strains SBFEM},\\
    &\boldsymbol{\kappa} = \mathbf{b}_{1}(\eta) \frac{\partial\mathbf{v}}{\partial\xi} + \frac{1}{\xi}\mathbf{b}_{2}(\eta) \frac{\partial\mathbf{v}}{\partial\eta} \label{eq: curvature strains SBFEM},\\
        &\boldsymbol{\gamma} = \mathbf{s}_{1}(\eta) \frac{\partial\mathbf{v}}{\partial\xi} + \frac{1}{\xi}\mathbf{s}_{2}(\eta) \frac{\partial\mathbf{v}}{\partial\eta} + \mathbf{s}_{3} \mathbf{v} \label{eq: shear strains SBFEM}.
\end{align} 
\end{subequations}
The SBFEM parametrization being established, the numerical approximation is introduced in the following section.
\section{Numerical approximation} \label{sec: Numerical approximation}
\subsection{Interpolation of the displacement field}
In the underlying formulation, linear Lagrangian interpolation functions are applied to derive a low-order polygonal SBFEM element formulation. Meanwhile, the interpolations in radial direction are independent of the interpolations in scaling direction. In a first step, the shape functions in circumferential direction are introduced. Here, the isoparametric concept is taken into account, which means that the same shape functions are applied for the discretization of the geometry and the displacement field. Recalling Eq.\eqref{eq: Position vector boundary node}, this implies 
\begin{equation}
    \mathbf{d}^{h} = \underbrace{\begin{bmatrix}
        \bar{N}_{1}\mathbf{I}_{n_{DOF}\times n_{DOF}} &  \bar{N}_{2}\mathbf{I}_{n_{DOF}\times n_{DOF}}
    \end{bmatrix}}_{\bar{\mathbf{N}}(\eta)}\bar{\mathbf{d}}(\xi),\label{eq: interpolation in boundary direction}
\end{equation}
where $n_{DOF}=5$ is the number of degrees of freedom per node and $\tilde{\mathbf{d}}$ is a vector containing the nodal displacements at the boundary in circumferential direction. 

Next, the discretization of the displacement field in scaling direction is introduced with
\begin{equation}
    \bar{\mathbf{d}}(\xi) = \underbrace{\begin{bmatrix}
        \hat{N}_{1}(\xi) \mathbf{I}_{2n_{DOF}\times 2n_{DOF}} & \hat{N}_{2}(\xi) \hat{\mathbf{I}}
    \end{bmatrix}}_{\hat{\mathbf{N}}(\xi)} \begin{bmatrix}
        \hat{\mathbf{d}}_{1} \\ \hat{\mathbf{d}}_{2} \\ \hat{\mathbf{d}}_{0}
    \end{bmatrix}, \label{eq: interpolation in radial direction}
\end{equation}
where $\hat{\mathbf{I}}$ is a matrix defined as $\hat{\mathbf{I}}^{\text{T}} = \begin{bmatrix}
    \mathbf{I}_{n_{DOF}\times n_{DOF}} \,\mathbf{I}_{n_{DOF}\times n_{DOF}}
\end{bmatrix}$ and $\hat{N}_{1}=\xi$ and $\hat{N}_{2}=1-\xi$ are the shape functions in radial direction. $\hat{\mathbf{d}}_{0}$ is the displacement vector of the scaling center, while $\hat{\mathbf{d}}_{i}$ are the displacement vectors of the boundary nodes.

Inserting Eq.\eqref{eq: interpolation in radial direction} into \eqref{eq: interpolation in boundary direction}, the approximation of the displacement field and its variation are given by
\begin{subequations}
    \begin{align}
        &\mathbf{d}^{h}= \bar{\mathbf{N}}(\eta)\hat{\mathbf{N}}(\xi)\hat{\mathbf{d}} \quad \text{and} \\ &\delta\mathbf{d}^{h}= \bar{\mathbf{N}}(\eta)\delta\hat{\mathbf{N}}(\xi)\hat{\mathbf{d}}. \label{eq: approximated displacement field}
    \end{align}
\end{subequations}
\subsection{Approximation of the strain operators}
The approximated displacement field from Eq.\eqref{eq: approximated displacement field} is introduced into Eqs.\eqref{eq: membrane strains SBFEM}, \eqref{eq: curvature strains SBFEM} and \eqref{eq: shear strains SBFEM}, to receive the approximated membrane, curvature and shear strains respectively. This leads to
\begin{equation}
\boldsymbol{\varepsilon}^{h} = \mathbf{B}_{m}\hat{\mathbf{u}}, \qquad \boldsymbol{\kappa}^{h} = \mathbf{B}_{b} \hat{\mathbf{v}}, \qquad \boldsymbol{\gamma}^{h} = \mathbf{B}_{s} \hat{\mathbf{v}}.
\end{equation}
Here, $\mathbf{B}_{m}$, $\mathbf{B}_{b}$ and $\mathbf{B}_{s}$ are the strain displacement operators defined as 
\begin{subequations}
    \begin{align}
          &\mathbf{B}_{m}= \mathbf{M}_{1}(\xi,\eta) + \frac{1}{\xi} \mathbf{M}_{2}(\xi,\eta), \\ & \mathbf{B}_{b}= \mathbf{B}_{1}(\xi,\eta) + \frac{1}{\xi} \mathbf{B}_{2}(\xi,\eta), \\ & \mathbf{B}_{s}= \mathbf{S}_{1}(\xi,\eta) + \frac{1}{\xi} \mathbf{S}_{2}(\xi,\eta)+\mathbf{S}_{3}(\xi,\eta), 
    \end{align}\label{eq. Bs}
\end{subequations}
with
\begin{subequations}
    \begin{align}
        &\mathbf{M}_{1}(\xi,\eta) = \mathbf{m}_{1}(\eta)\bar{\mathbf{N}}(\eta)\hat{\mathbf{N}}(\xi)_{,\xi}, \\ &\mathbf{M}_{2}(\xi,\eta) = \mathbf{m}_{2}(\eta)\bar{\mathbf{N}}(\eta)_{,\eta}\hat{\mathbf{N}}(\xi), \\
                &\mathbf{B}_{1}(\xi,\eta) = \mathbf{b}_{1}(\eta)\bar{\mathbf{N}}(\eta)\hat{\mathbf{N}}(\xi)_{,\xi}, \\ & \mathbf{B}_{2}(\xi,\eta) = \mathbf{b}_{2}(\eta)\bar{\mathbf{N}}(\eta)_{,\eta}\hat{\mathbf{N}}(\xi),  \\
               & \mathbf{S}_{1}(\xi,\eta) = \mathbf{s}_{1}(\eta)\bar{\mathbf{N}}(\eta)\hat{\mathbf{N}}(\xi)_{,\xi}, \\& \mathbf{S}_{2}(\xi,\eta) = \mathbf{s}_{2}(\eta)\bar{\mathbf{N}}(\eta)_{,\eta}\hat{\mathbf{N}}(\xi), \\ &
        \mathbf{S}_{3}(\xi,\eta) = \mathbf{s}_{3}\bar{\mathbf{N}}(\eta)\hat{\mathbf{N}}(\xi).
    \end{align}
\end{subequations}  
$\boldsymbol{\varepsilon}$, $\boldsymbol{\kappa}$ and $\boldsymbol{\gamma}$ can now be summarized in the global strain field, such that
\begin{equation}
    \boldsymbol{\epsilon}^{h}_{g} (\mathbf{d}) = \mathbf{B} \hat{\mathbf{d}} \quad \text{and} \quad \delta\boldsymbol{\epsilon}^{h}_{g} (\mathbf{d}) = \mathbf{B} \delta\hat{\mathbf{d}}, \label{eq: e = Bd}
\end{equation}
where 
\begin{multline}
    \mathbf{B} = \begin{bmatrix}
        \mathbf{M}_{1}(\xi,\eta) & \mathbf{0}\\
        \mathbf{0} & \mathbf{B}_{1}(\xi,\eta)\\
        \mathbf{0} & \mathbf{S}_{1}(\xi,\eta)
    \end{bmatrix} \\+ \frac{1}{\xi}\begin{bmatrix}
        \mathbf{M}_{2}(\xi,\eta) & \mathbf{0}\\
        \mathbf{0} & \mathbf{B}_{2}(\xi,\eta)\\
        \mathbf{0} & \mathbf{S}_{2}(\xi,\eta)
    \end{bmatrix}+\begin{bmatrix}
        \mathbf{0} & \mathbf{0}\\ \mathbf{0} & \mathbf{0} \\ \mathbf{0} & \mathbf{S}_{3}(\xi,\eta)
    \end{bmatrix}.
\end{multline}

\subsection{Element stiffness matrix}
The approximated displacement and strain fields can now be introduced into the weak form of equilibrium in Eq.\eqref{eq: Weak form}, yielding
\begin{multline}
    \delta \hat{\mathbf{d}}\Big(\int_{\mathit{\Omega}_{s}}(\mathbf{B})^{\text{T}}\mathbf{C}\mathbf{B}\,\text{d}A \hat{\mathbf{d}}- \int_{\mathit{\Omega}_{s}}(\bar{\mathbf{N}}(\eta)\hat{\mathbf{N}}(\xi))^{\text{T}} \mathbf{b}\,\text{d}A\\-\int_{\partial\mathit{\Omega}_{s}}(\bar{\mathbf{N}}(\eta)\hat{\mathbf{N}}(\xi))^{\text{T}}\bar{\mathbf{t}}\,\text{d}S\Big)
\end{multline}
at section level. In this case, the sectional stiffness matrix $\mathbf{K}_{sec}$ is defined as $\mathbf{K}_{sec} = \int_{\mathit{\Omega}_{s}}\mathbf{B}^{\text{T}}\mathbf{C}\mathbf{B}\,\text{d}A$. The sectional stiffness matrix can be split into the bending and shear stiffness in the same manner as the strain displacement operator, such that
\begin{subequations}
   \begin{align}
       &\mathbf{K}_{m,sec} = \int_{\mathit{\Omega}_{s}} \mathbf{B}_{m}^{\text{T}}\mathbf{C}_{m}\mathbf{B}_{m}\text{dA}, \\ &\mathbf{K}_{b,sec} = \int_{\mathit{\Omega}_{s}} \mathbf{B}_{b}^{\text{T}}\mathbf{C}_{b}\mathbf{B}_{b}\text{dA}, \\ &\mathbf{K}_{s,sec} = \int_{\mathit{\Omega}_{s}} \mathbf{B}_{s}^{\text{T}}\mathbf{C}_{s}\mathbf{B}_{s}\text{dA} .\label{eq: Ks}
   \end{align} \label{eq: shear stiffness matrix}
\end{subequations}
The sectional stiffness matrices are then assembled to derive the element stiffness matrix with
\begin{equation}
    \mathbf{K}_{e} = \vcenter{\hbox{\LARGE\textbf{A}}}_{i=1}^{n_{s}} \mathbf{K}_{sec,i} .
\end{equation}
However, the SBFEM formulation with the shear stiffness matrix from Eq.\eqref{eq: shear stiffness matrix} locks in the thin plate limit, as shown in Tab.\ref{tab: 4}, due to an overestimation of the shear stiffness if a locking remedy is lacking. Therefore, an assumed natural strain field will be introduced in the following section.
\section{Assumed natural strain field}\label{sec: Assumed natural strain field}
Low-order Reissner--Mindlin element formulations suffer from transverse shear locking in the thin plate limit. A crucial step in the derivation of a Reissner--Mindlin plate formulation is therefore the derivation of a locking remedy. A possible approach are the assumed natural strains (ANS) originally proposed in \cite{Bathe1985} for a bilinear quadrilateral finite plate element. However, the approach highly depends on the element shape and a direct application on polygonal elements is not possible. In this section, an assumed natural strain field is introduced at section level, allowing to derive an assumed natural strain approach for polygonal SBFEM elements. 
\subsection{Assumed shear strain field}
\begin{figure}
    \centering
    \input{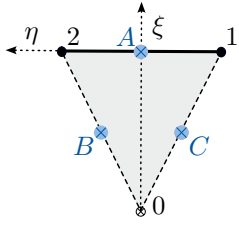}
    \caption{Positioning of the tying points $A$, $B$ and $C$ for the assumed natural strain method against transverse shear locking in a scaled boundary finite element section.}
    \label{fig: 3}
\end{figure}

In the ANS-approach, a distinctive interpolation of the shear strains is introduced. This means that the shear strains are no longer evaluated at the nodes, but at the so-called tying points. In case of a triangular, bilinear SBFEM section, three tying points are introced. The first tying point, point $A$ is located at midspan between the boundary nodes. The other two points, $B$ and $C$ are located at midspan between one boundary node and the scaling center. Their coordinates with respect to the parametric space are given in Tab.\ref{tab:Parametric coordinates of the tying points in the SBFEM section.} and they are illustrated in Fig.\ref{fig: 3}.
\begin{table}[htbp]
\caption{Parametric coordinates of the tying points in the SBFEM section.} \label{tab:Parametric coordinates of the tying points in the SBFEM section.}%
\begin{tabular}{@{}llll@{}}
\toprule
Point & A & B & C\\
\midrule
$\xi$-coordinate & $1$ &$1/2$&$1/2$ \\
$\eta$-coordinate &  $0$ & $1$&$-1$ \\
\botrule
\end{tabular}
\end{table}

The shear strains are linearly interpolated using the shape functions of a triangular bilinear SBFEM section, which leads to
\begin{subequations}
    \begin{align}
         &\Tilde{\gamma}_{\xi z} = \frac{1}{2}(1+\eta) \Tilde{\gamma}_{\xi z}^{B} + \frac{1}{2}(1-\eta)\Tilde{\gamma}_{\xi z}^{C}, \\ & \Tilde{\gamma}_{\eta z} = \xi \Tilde{\gamma}_{\eta z}^{A} .
    \label{eq: interpolated strains}
    \end{align}
\end{subequations}
The local shear strains are derived from $\tilde{\gamma}_{\xi z}^{N} = w_{,\xi} + \beta_{\xi}$, with $N=\{B,C\}$ and $\tilde{\gamma}_{\eta z}^{A} = w_{,\eta} + \beta_{\eta}$ by replacing the rotations in the parametric space by the rotations in the physical space using 
\begin{equation}
    \begin{bmatrix}
    \beta_{\xi} \\ \beta_{\eta}
    \end{bmatrix} = \mathbf{J} \begin{bmatrix}
    \beta_{x} \\ -\beta_{y}
    \end{bmatrix}.
\end{equation}
This yields
\begin{subequations}
    \begin{align}
    &\Tilde{\gamma}_{\xi z}^{N} = \frac{\partial w}{\partial \xi} + \beta_{x} \frac{\partial x}{\partial \xi} - \beta_{y} \frac{\partial y}{\partial \xi}, \quad \text{and} \\ & \Tilde{\gamma}_{\eta z}^{A} = \frac{\partial w}{\partial \eta} + \beta_{x} \frac{\partial x}{\partial \eta} - \beta_{y} \frac{\partial y}{\partial \eta}.
    \end{align}
    \label{Eq: local shear strains}
\end{subequations}
The rotations at the tying points result from the interpolation of the nodal rotations using the shape functions of the scaled boundary section, defined as $N_{0}=1-\xi$, $N_{1} = \frac{1}{2}\xi(1-\eta)$ and $N_{2} = \frac{1}{2}\xi(1+\eta)$. They can be derived to
\begin{subequations}
\begin{align}
         &\beta_{x}^{A} = \frac{1}{2} (\theta_{y}^1 + \theta_{y}^2),\qquad \beta_{y}^{A} = \frac{1}{2} (\theta_{x}^1 + \theta_{x}^2),  \\
         &\beta_{x}^{B} =\frac{1}{2}  (\theta_{y}^2 + \theta_{y}^0),\qquad \beta_{y}^{B} = \frac{1}{2}  (\theta_{x}^2 + \theta_{x}^0) ,\\
         & \beta_{x}^{C} = \frac{1}{2}  (\theta_{y}^1 + \theta_{y}^0), \qquad \beta_{y}^{C} = \frac{1}{2}  (\theta_{x}^1 + \theta_{x}^0).
    \end{align}
      \label{eq: rotations at tying points}
\end{subequations}
Using the partial derivatives of the shape functions with respect to $\xi$ and $\eta$, the partial derivatives of the coordinates and the displacements result at the tying points can be deduced and are given in Tab.\ref{tab:2}.
\begin{table}[htbp]
\caption{Partial derivatives at the tying points $A$, $B$ and $C$.} \label{tab:2}%
\begin{tabular}{@{}lccc@{}}
\toprule
& \multicolumn{1}{@{}c@{}}{$/\partial \eta$} & \multicolumn{2}{@{}c@{}}{$/\partial \xi$} \\\cmidrule(lr{0.5em}){2-2} \cmidrule(lr{0.5em}){3-4}
 & A & B & C\\
\midrule
$\partial x ^{i}$ & $\frac{1}{2}(x_{2}-x_{1})$ & $ x_{2}-x_{0}$&  $ x_{1}-x_{0}$\\
 $\partial y ^{i}$  &$\frac{1}{2}(y_{2}-y_{1})$  & $y_{2}-y_{0}$ &$y_{1}-y_{0}$\\
 $\partial w ^{i}$ &$\frac{1}{2}(w_{2}-w_{1})$ &  $w_{2}-w_{0}$ & $w_{1}-w_{0}$\\
\botrule
\end{tabular}
\end{table}

Introducing the results of Eqs.\eqref{eq: rotations at tying points} and the partial derivatives in Tab.\ref{tab:2} into Eq.\eqref{Eq: local shear strains}, yields the local shear strains evaluated at the tying points $A$, $B$ and $C$.
\begin{subequations}
    \begin{multline}
        \Tilde{\gamma}_{\eta z}^{A} = 
        \frac{1}{2}(w_{2}-w_{1})+ \frac{1}{4}(x_{2}-x_{2})(\theta_{y}^{1}+\theta_{y}^{2})\\-\frac{1}{4}(y_{2}-y_{1})(\theta_{x}^{1}+\theta_{x}^{2})
    \end{multline}
    \begin{multline}
        \Tilde{\gamma}_{\xi z}^{B}=
        (w_{2}-w_{0})+\frac{1}{2}(x_{2}-x_{0})(\theta_{y}^{2}+\theta_{y}^{0})\\-\frac{1}{2}(y_{2}-y_{0})(\theta_{x}^{2}+\theta_{x}^{0})
    \end{multline}
    \begin{multline}
        \Tilde{\gamma}_{\xi z}^{C}= 
        (w_{1}-w_{0}) + \frac{1}{2}(x_{1}-x_{0})(\theta_{y}^{1}+\theta_{y}^{0})\\-\frac{1}{2}(y_{1}-y_{0})(\theta_{x}^{1}+\theta_{x}^{0})
    \end{multline}
\end{subequations} 
The local shear strains at the tying points are now inserted into Eqs.\eqref{eq: interpolated strains}, implying
\begin{subequations}
    \begin{multline}
        \Tilde{\gamma}_{\xi z} = \frac{1}{2}(1+\eta) \Big[ (w_{2}-w_{0})+\frac{1}{2}(x_{2}-x_{0})(\theta_{y}^{2}+\theta_{y}^{0})\\-\frac{1}{2}(y_{2}-y_{0})(\theta_{x}^{2}+\theta_{x}^{0})\Big]\\ + \frac{1}{2}(1-\eta)\Big[(w_{1}-w_{0}) + \frac{1}{2}(x_{1}-x_{0})(\theta_{y}^{1}+\theta_{y}^{0})\\-\frac{1}{2}(y_{1}-y_{0})(\theta_{x}^{1}+\theta_{x}^{0})],
    \end{multline}
    \begin{multline}
         \Tilde{\gamma}_{\eta z} = \xi \Big[\frac{1}{2}(w_{2}-w_{1})+ \frac{1}{4}(x_{2}-x_{1})(\theta_{y}^{1}+\theta_{y}^{2})\\-\frac{1}{4}(y_{2}-y_{1})(\theta_{x}^{1}+\theta_{x}^{2})\Big].
    \end{multline}
\end{subequations}
The last step consists of the transformation of the local strains to global strains using the inverse of the Jacobian matrix $\mathbf{J}$
\begin{equation}
    \boldsymbol{\gamma}=\begin{bmatrix}
        \gamma_{xz} \\ \gamma_{yz}
    \end{bmatrix} = \mathbf{J}^{-1}\begin{bmatrix}
        \tilde{\gamma}_{\xi z} \\ \tilde{\gamma}_{\eta z}
    \end{bmatrix} = \mathbf{J}^{-1} \tilde{\boldsymbol{\gamma}}.
\end{equation}
\subsection{B-operator of the shear strains}
The strain displacement operator for the shear strains can now be reformulated as
\begin{multline}
    \mathbf{B}_{m}^{ANS} =  \mathbf{J}^{-1} \left[ \begin{array}{ccc|ccc}
    a_{11} & a_{12} & a_{13} \\
    a_{21} & a_{22} & a_{23} \\
    \end{array}  \right. \\  \left. \begin{array}{ccc|ccc} 
    b_{11} & b_{12} & b_{13}&c_{11} & c_{12} & c_{13} \\
     b_{21} & b_{22} & b_{23}&c_{21} & c_{22} & c_{23}
    \end{array} \right],
\end{multline}
with
\begin{align}
        a_{11} &= \frac{1}{2}(1-\eta) &  a_{21} &= -\frac{1}{2}\xi \nonumber \\
         a_{12} &= \frac{1}{4}(1-\eta)(x_{1}-x_{0}) & a_{22} &= \frac{1}{4}\xi(x_{2}-x_{1}) \nonumber\\
         a_{13} &= -\frac{1}{4}(1-\eta)(y_{1} - y_{0}) & a_{23} &= -\frac{1}{4}\xi(y_{2}-y_{1})
    \end{align}
\begin{align}
        b_{11} &= \frac{1}{2}(1+\eta) &  b_{21} &= \frac{1}{2}\xi \nonumber\\
        b_{12}&=\frac{1}{4}(1+\eta)(x_{2}- x_{0}) & b_{22} &= \frac{1}{4}\xi(x_{2}-x_{1}) \nonumber \\
       b_{13} &= -\frac{1}{4}(1+\eta)(y_{2}-y_{0} ) & b_{23} &= -\frac{1}{4}\xi(y_{2}-y_{1})
    \end{align}
an
    \begin{align}
        c_{11} &= -1 &  c_{21} &= 0 \nonumber\\
        c_{12} &= \frac{1}{4}\Big[x_{2}+\eta x_{2}+ x_{1}-\eta x_{1} -2x_{0}\Big] & c_{22} &= 0 \nonumber\\
       c_{13} &= -\frac{1}{4}\Big[ y_{2}+\eta y_{2}- y_{1}+\eta y_{1} -2\eta x_{0}\Big] & c_{23} &= 0.
    \end{align}

To consider the ANS approach in the SBFEM formulation, the $\mathbf{B}_{s}$-operator for the transverse shear part in Eq.\eqref{eq. Bs} has to be replaced by the newly derived $\mathbf{B}_{s}^{ANS}$ operator. This yields the sectional shear stiffness matrix $\mathbf{K}_{s,sec}^{ANS}$ written as

\begin{equation}
        \mathbf{K}_{s,sec}^{ANS} = \int_{\mathit{\Omega}_{s}} (\mathbf{B}_{s}^{ANS})^{\text{T}}\mathbf{C}_{s}\mathbf{B}_{s}^{ANS}\text{dA},
\end{equation}
which replaces the original shear stiffness matrix from Eq.\eqref{eq: Ks}.

In the upfollowing numerical examples, the formulation including the ANS approach will be distinguished from the standard SBFEM formulation by the subscript $(\cdot)_{ANS}$.
\section{Incorporation of three dimensional material laws} \label{sec: Incorporation of 3D material laws}
The plate formulation can be generalized by the incorporation of three dimensional constitutive laws, where the plane stress assumptions are no longer enforced in the strong form of the material laws, but are weakly fulfilled. The following derivations are based on \cite{Klinkel2008} and limited to linear material models.
\subsection{Three dimensional constitutive equations}
In this work, isotropic and linear elastic material laws are considered. This means that the stresses depend linearly on the strains, such that
\begin{equation}
    \boldsymbol{\sigma} = \mathbb{C}\boldsymbol{\epsilon}\label{eq: S=CE}, 
\end{equation}
where $\boldsymbol{\sigma} =  \begin{bmatrix}
        \sigma_{11} &\sigma_{22} & \sigma_{33} & \sigma_{12} & \sigma_{13} & \sigma_{23}
    \end{bmatrix}^{\text{T}}$ and $\boldsymbol{\epsilon} = \begin{bmatrix}
          \epsilon_{11} &  \epsilon_{22} &  \epsilon_{33} & 2\epsilon_{12} & 2\epsilon_{13}  &  2\epsilon_{23}
    \end{bmatrix}^{\text{T}} $.  $\mathbb{C}$ is the material matrix containing the Lamé constants $\lambda$ and $\mu$ defined as
\begin{equation}
    \mathbb{C} =  \begin{bmatrix}
        \lambda+2\mu & \lambda & \lambda & 0 & 0& 0\\
        \lambda & \lambda + 2\mu & \lambda & 0 & 0 & 0\\
        \lambda & \lambda & \lambda + 2\mu & 0 & 0 & 0\\
        &&&\mu & 0& 0\\
        & \text{sym.}& & & \mu & 0\\
        & & & & & \mu
    \end{bmatrix}.
\end{equation}
 These constants can be expressed in dependence of the the Young's modulus $E$ and the Poisson's ratio $\nu$ by
 \begin{equation}
     \lambda = \frac{E\nu}{(1+\nu)(1-2\nu)}, \quad \mu = \frac{E}{2(1+\nu)}.
 \end{equation}
\subsection{Derivation of the strain vector}
The 3D strain in vector representation using Voigt notation reads $\boldsymbol{\epsilon}$. It depends on the geometrical strains $\boldsymbol{\epsilon}_{g}$ from Eq.\eqref{eq: geometrical strains} and the thickness strains $\boldsymbol{\epsilon}_{z}$. It is formulated as 
\begin{equation}
    \boldsymbol{\epsilon} = \mathbf{A}\begin{bmatrix}
        \boldsymbol{\epsilon}_{g} \\ \boldsymbol{\epsilon}_{z}
    \end{bmatrix}\label{eq: E = Ae}.
\end{equation}
As in \cite{Klinkel2008}, the thickness strain vector contains two independent parameters and is given by 
\begin{equation}
    \boldsymbol{\epsilon}_{z} = \begin{bmatrix}
    \varepsilon_{zz}^{0} & \varepsilon_{zz}^{1}
\end{bmatrix}^{\text{T}}.
\end{equation}
Meanwhile, the transformation matrix $\mathbf{A} = \begin{bmatrix}
    \mathbf{A}_{1},\mathbf{A}_{2}
\end{bmatrix}$ contains the submatrices $\mathbf{A}_{i}$ defined as
\begin{equation}
    \mathbf{A}_{1} = \begin{bmatrix}
        1&0&0&\zeta&0&0&0&0\\
         0&1&0&0&\zeta&0&0&0\\
          0&0&0&0&0&0&0&0\\
           0&0&1&0&0&\zeta&0&0\\
           0&0&0&0&0&0&1&0\\
           0&0&0&0&0&0&0&1
    \end{bmatrix}, \quad
        \mathbf{A}_{2} = \begin{bmatrix}
            0 & 0 \\
            0 & 0\\
            1 & \zeta \\
            0 & 0\\
            0 & 0\\
            0 & 0
    \end{bmatrix}. \label{eq: A1 and A2}
\end{equation}
It is important to note that a linear interpolation of the thickness strains in thickness direction ($\zeta$), as in $\mathbf{A}_{2}$, is necessary to overcome the effect of Poisson's thickness locking, which occurs in case of non-zero Poisson's ratio and three dimensional material laws. This will be visualized in Example \ref{sec: Poisson's thickness locking analysis} of Section \ref{sec: Numerical examples}, where results are compared for $\mathbf{A}_{2}$ from Eq.\eqref{eq: A1 and A2} and $\mathbf{A}_{2}^{*} = \begin{bmatrix}
    0\,0\,1\,0\,0\,0
\end{bmatrix}^{\text{T}}$.

Next, the weak form of the formulation is presented.
\subsection{Weak form of the two field variational formulation}
In the three dimensional case, the weak form is written as
\begin{multline}\label{eq: 3D variational formulation}
\int_{B}\delta\boldsymbol{\epsilon}^{\text{T}}\boldsymbol{\sigma}\,\text{d}V - \int_{ \mathit{\Omega}} \delta \mathbf{d}^{\text{T}}\mathbf{b}\,\text{d}A \\- \int_{\partial \mathit{\Omega}}\delta \mathbf{d}^{\text{T}}\bar{\mathbf{t}}\,\text{d}S = 0.
\end{multline}
By introducing Eq.\eqref{eq: S=CE} and Eq.\eqref{eq: E = Ae}, the variational formulation can be rewritten as
\begin{multline}
\int_{\mathit{\Omega}} \begin{bmatrix}
        \delta\boldsymbol{\epsilon}_{g} & \delta\boldsymbol{\epsilon}_{z}
    \end{bmatrix} \int_{t} \mathbf{A}^{\text{T}}\mathbb{C}\mathbf{A} \,\text{d}\zeta \begin{bmatrix}
        \boldsymbol{\epsilon}_{g} \\ \boldsymbol{\epsilon}_{z}
    \end{bmatrix}\, \text{d}A \\  - \int_{\mathit{\Omega}} \delta \mathbf{d}^{\text{T}}\mathbf{b}\,\text{d}A - \int_{\partial \mathit{\Omega}}\delta \mathbf{d}^{\text{T}}\bar{\mathbf{t}}\,\text{d}S= 0.
\end{multline}
Since the thickness strains are independent of the displacements, it is possible to derive a two field variational formulation equivalent to
\begin{subequations}
    \begin{multline}
        \int_{\mathit{\Omega}}\bigg[ \delta\boldsymbol{\epsilon}_{g}^{\text{T}}\int_{t} \mathbf{A}_{1}^{\text{T}} \mathbb{C}\mathbf{A}_{1}\,\text{d}\zeta \boldsymbol{\epsilon}_{g} \\+ \delta\boldsymbol{\epsilon}_{g}^{\text{T}}\int_{t} \mathbf{A}_{1}^{\text{T}} \mathbb{C}\mathbf{A}_{2}\,\text{d}\zeta \boldsymbol{\epsilon}_{z} \bigg]\text{d}A \\-\int_{\mathit{\Omega}} \delta\mathbf{d}^{\text{T}} \mathbf{b} \, \text{d}A-    \int_{\partial \mathit{\Omega}} \delta\mathbf{d}^{\text{T}} \mathbf{t} \, \text{d}S = 0
    \end{multline}
    \begin{multline}
\int_{\mathit{\Omega}}\bigg[\delta\boldsymbol{\epsilon}_{z}^{\text{T}} \int_{t}\mathbf{A}_{2}^{\text{T}} \mathbb{C}\mathbf{A}_{1}\, \text{d}\zeta \boldsymbol{\epsilon}_{g} \\+ \delta\boldsymbol{\epsilon}_{z}^{\text{T}} \int_{t}\mathbf{A}_{2}^{\text{T}} \mathbb{C}\mathbf{A}_{2} \, \text{d}\zeta\boldsymbol{\epsilon}_{z}  \bigg] \text{d}A = 0.
    \end{multline}\label{Twovariational formulation}
\end{subequations}
It allows a different parameterization of the variational fields $\delta \boldsymbol{\epsilon}_{g}$ and $\delta\boldsymbol{\epsilon}_{z}$. Simplifiying the derivations in \cite{Klinkel2008}, the thickness strain parameters are assumed to be constant over section domain and the geometrical strains are given by Eq.\eqref{eq: geometrical strains}
\subsection{Extended element stiffness matrix}
Introducing the parameterizations in the variational formulations in Eq.\eqref{Twovariational formulation} yields the following weak forms
\begin{subequations}
    \begin{multline}
        \delta\hat{\mathbf{d}}^{\text{T}}\Big(\mathbf{K}_{dd}\hat{\mathbf{d}} + \mathbf{K}_{dz}\boldsymbol{\epsilon}_{z}\Big) -\delta \hat{\mathbf{d}}^{\text{T}}\int_{\mathit{\Omega}_{s}} \hat{\mathbf{N}}^{\text{T}}\bar{\mathbf{N}}^{\text{T}} \mathbf{b} \, \text{d}A \\ -    \delta\hat{\mathbf{d}}^{\text{T}}\int_{\partial \mathit{\Omega}_{s}} \hat{\mathbf{N}}^{\text{T}}\bar{\mathbf{N}}^{\text{T}} \mathbf{t} \, \text{d}S,\label{eq: variation over d}
    \end{multline}
    \begin{equation} \delta\boldsymbol{\epsilon}_{z}^{\text{T}}\Big(\mathbf{K}_{zd} \hat{\mathbf{d}} + \mathbf{K}_{zz} \boldsymbol{\epsilon}_{z}\Big) = 0.\label{eq: variation over beta}
\end{equation}
\end{subequations}
Here, the stiffness matrices $\mathbf{K}_{dd}$, $ \mathbf{K}_{zz}$, $\mathbf{K}_{dz}$ and $\mathbf{K}_{zd}$ are defined as
\begin{subequations}
    \begin{align}
        &\mathbf{K}_{dd} = \int_{\mathit{\Omega}_{s}}\mathbf{B}^{\text{T}}\int_{t}\mathbf{A}_{1}^{\text{T}} \mathbb{C}\mathbf{A}_{1} \, \text{d}\zeta \mathbf{B} \, \text{d}A, \\
        &\mathbf{K}_{zz} = \int_{\mathit{\Omega}_{s}}\int_{t}\mathbf{A}_{2}^{\text{T}} \mathbb{C}\mathbf{A}_{2} \, \text{d}\zeta \, \text{d}A,\\
        &\mathbf{K}_{zd} = \int_{\mathit{\Omega}_{s}}\int_{t}\mathbf{A}_{2}^{\text{T}} \mathbb{C}\mathbf{A}_{1} \, \text{d}\zeta \mathbf{B} \, \text{d}A = \mathbf{K}_{dz}^{\text{T}}.
    \end{align}
\end{subequations}
Note that in the linear case, an analytical integration in thickness direction is possible. 

The zero stress condition in thickness direction is enforced at section level by Eq.\eqref{eq: variation over beta}, which can be reformulated as
\begin{equation}
    \boldsymbol{\epsilon}_{z} = - \mathbf{K}_{zz}^{-1}\mathbf{K}_{zd}\hat{\mathbf{d}}.
\end{equation}
Introducing the results in Eq.\eqref{eq: variation over d}, yields the section stiffness matrix for three dimensional material laws defined as
\begin{equation}
    \mathbf{K}_{sec}^{\text{3D}} = \mathbf{K}_{dd} - \mathbf{K}_{dz}\mathbf{K}_{zz}^{-1}\mathbf{K}_{zd}.
\end{equation}

\section{Numerical examples}\label{sec: Numerical examples}
The element formulation is evaluated in multiple numerical examples proving its stability and applicability on various mesh types, while highlighting the effectiveness of the ANS approach from Section \ref{sec: Assumed natural strain field}. The geometries analyzed using are a cantilever plate, a clamped/simply-supported square plate, a clamped circular plate and a L-bracket. Here different load cases are applied. The element formulations tested or considered as references are described below

\begin{itemize}
  \item SBFEM: fully discretized scaled boundary finite plate element with linear shape functions in scaling and circumferential direction without locking remedy.
  \item SBFEM$_{R}$: fully discretized scaled boundary finite plate element with linear shape functions in scaling and circumferential direction with reduced integration.
  \item SBFEM$_{SR}$: fully discretized scaled boundary finite element with linear shape functions in scaling and circumferential direction with reduced integration of the shear stiffness part (selective reduced).
  \item SBFEM$_{ANS}$: fully discretized scaled boundary finite plate element with linear shape functions and assumed natural strains, see Section \ref{sec: Assumed natural strain field}.
  \item SBFEM$_{ANS}\mathbf{A}_2$: fully discretized scaled boundary finite element including membrane strains and linearly interpolated thickness strains, see Section \ref{sec: Incorporation of 3D material laws}. 
  \item SBFEM$_{ANS}\mathbf{A}_{2}^{*}$: fully discretized scaled boundary finite element including membrane strains and a constant  interpolation of thickness strains, see Section \ref{sec: Incorporation of 3D material laws}. 
  \item Q1$_{ANS}$: bilinear quadrilateral finite plate element with assumed natural strains, see \cite{Bathe1985}.
  \item MT3: bilinear triangular finite element with assumed natural strains, see \cite{Pal-Gap1999}.
  \item PRMn-W: Polygonal Reissner--Mindlin plate element based on Wachspress shape functions, see \cite{Nguyen2017}.
  \item PRMn-PL: Polygonal Reissner--Mindlin plate element based on piecewise-linear shape functions, see \cite{Nguyen2017}.
  \item PRMn-T3: Triangular Reissner--Mindlin plate element with linear shape functions, see \cite{Nguyen2017}.
  \item PRMn-Q4: Four-node quadrilateral Reissner--Mindlin plate element based on barycentric shape functions, see \cite{Nguyen2017}.
\end{itemize}
\subsection{Zero-energy mode test}
\begin{figure*}
    \centering
    \includegraphics[width=\linewidth]{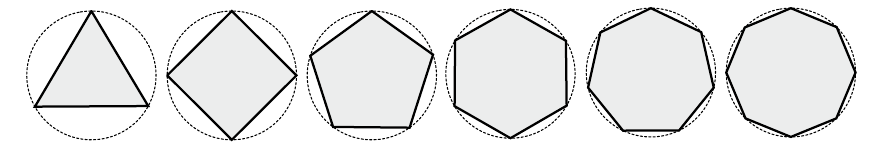}
    \caption{Element shapes used for the zero-energy mode test.}
    \label{fig: 4}
\end{figure*}

First a zero-energy mode test, also known as ellipticity condition \cite{Bathe2011}, is carried out to prove the stability of the SBFEM$_{ANS}$ formulation, where each node has three degrees of freedom being $w$, $\beta_{x}$ and $\beta_{y}$. Therefore, a set of 6 unsupported element models is evaluated, see Fig.\ref{fig: 4}. As in \cite{Nguyen2023}, the material parameters are given by Young's modulus $E =10^5$ and Poisson's ratio $\nu=0.25$. The thickness $t$ is set to $t=0.01$. For each of the elements, the stiffness matrix is derived and their eigenvalues are calculated. Note that for the analysis of the eigenvalues of the stiffness matrices, a static condensation of the scaling centers is carried out. The results for the eigenvalues can be seen in Tab.\ref{tab: 3} As displayed, each of the elements possesses only three zero-energy modes, which correspond to the number of rigid-body modes. No spurious zero-energy modes occur and the formulation passes the test.
\begin{table*}[htbp]
\caption{Eigenvalues of the stiffness matrices of 6 unsupported polygonal SBFEM$_{ANS}$ elements.}\label{tab: 3}
\begin{tabular*}{\textwidth}{@{\extracolsep\fill}lcccccc}
\toprule%
& \multicolumn{6}{@{}c@{}}{Element type} \\\cmidrule{2-7}%
Eigenvalue & Triangle & Quadrangle&  Pentagon& Hexagon & Heptagon & Octagon\\
\midrule
1&0           & 0           & 0           & 0  & 0   & 0    \\
        2&0           & 0           & 0            & 0           & 0           & 0        \\
        3&0           & 0           & 0          &  0           & 0           & 0 \\
        4&0.00532938  & 0.005333316 & 0.005635866 & 0.005635866  & 0.00463307   & 0.004190258  \\
        5&0.00532938  & 0.006666667 & 0.005635866 & 0.005635866  & 0.00463307   & 0.004190258  \\
        6&0.009622504 & 0.011111111 & 0.010567295 & 0.010567295  & 0.008687016  & 0.007856742  \\
        7&72.16878365 & 0.017773228 & 0.028967636 & 0.028967636  & 0.011967864  & 0.011550872  \\
        8&351.8576548 & 0.017773228 & 0.028967636 & 0.028967636  & 0.011967864  & 0.011550872  \\
        9&351.8576548 & 111.1111111 & 4.454720799 & 4.454720799  & 0.032506667  & 0.017284832  \\
       10& -           & 312.54124263    & 4.454720799 & 4.454720799  & 0.032506667  & 0.03106814   \\
       11&  -         & 312.54124263  & 122.0003144 & 122.0003144  & 6.029733787  & 0.03106814   \\
       12& -          & 416.66800002 & 282.0703267 & 282.0703267  & 6.029733787  & 5.640569253  \\
      13&  -        & -           & 282.0703267 & 282.0703267  & 16.92487955  & 8.573476714  \\
       14& -           & -         & 502.3724971 & 502.3724971  & 16.92487955  & 8.573476714  \\
     15&   -      & -       & 502.3724971 & 502.3724971  & 113.9514956  & 20.39583821  \\
       16& -           & - & -           &-            & 236.6445896  & 20.39583821  \\
        17&-& -       & -           & -            & 236.6445896  & 106.3451979  \\
       18&-    & -     & -           & ~            & 546.6598292  & 217.4095387  \\
       19& -  & - & - & - & 546.6598292  & 217.4095387  \\
        20&-& - & - & - & 695.7172689  & 532.9592068  \\
        21& - & - & - & - & 695.7172689  & 532.9592068  \\
        22&-         & -& - & - & -&38.8164681  \\
        23&-     & -           & -           & - &-            & 738.8164681  \\
        24&- & - & - & - & - & 822.1185775  \\
\botrule
\end{tabular*}
\end{table*}
\subsection{Cantilever plate}
\begin{figure*}
    \centering
    \input{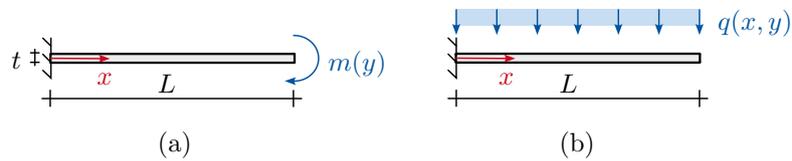}
    \caption{Illustration of the load cases (a) moment loading $m$ and (b) uniformly distributed load $q$ acting on the cantilever plate.}
    \label{fig: 5}
\end{figure*}

A cantilever plate of length $L=2$, width $B=1$ and a uniform thickness $t$ is first loaded by a moment line load $m(y)=1$ at the free edge and then by a uniformly distributed load $q(x,y)=-10$ in $z$-direction. The material parameters are given by Young's modulus $E=10^5$ and Poisson's ratio $\nu=0$. The problem set ups are visualized in Fig.\ref{fig: 5}.

During the evaluation process, the formulation is tested for relatively 'thick' plates and in the thin plate limit by setting the thickness to $t=\{1,\,0.1,\,0.01,\, 0.001\}$. The numerical results for the displacements $w$ at the free edge are compared to an analytical reference solution from Timoshenko's beam theory, given in Section \ref{sec: Moment loading} and Section \ref{sec: unifromly distributed load} for the different load cases.
\subsubsection{Moment loading}\label{sec: Moment loading}
First, the moment loading at the free edge is applied. In this case, the analytical solution for the displacements $w_{ref}$ at the free edge are calculated from
\begin{equation}
    w_{ref} = - \frac{mL^{2}}{2EI}, \qquad \text{with} \quad I=\frac{Bt^{3}}{12}. \label{eq: Ref. solution moment loading}
\end{equation}
The cantilever plate is discretized using only one quadrilateral SBFEM or SBFEM$_{ANS}$ element. The resulting relative errors $(w-w_{ref})/w_{ref}$ are displayed in Tab.\ref{tab: 4} for different plate thicknesses $t$. It can be observed that the error of the fully discretized SBFEM formulation is quite high and increases with the slenderness of the plate, displaying the typical effect of transverse shear locking in low-order Reissner--Mindlin formulations. By contrast, the incorporation of the assumed natural strain approach from Section \ref{sec: Assumed natural strain field} significantly improves the results, reducing the relative error of the SBFEM$_{ANS}$ formulation to zero.
\begin{figure}
\centering
    \input{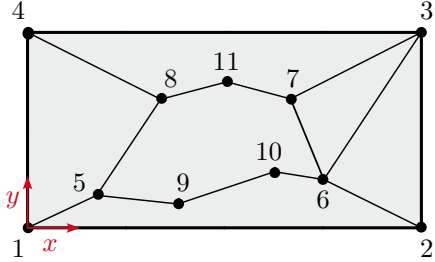}
    \caption{Mesh of the cantilever plate discretized using 6 polygonal elements.}
    \label{fig: 6}
\end{figure}
\begin{table}[htbp]
\caption{Evaluation of the tip displacemet $w$ of the cantilever plate loaded by $m(y)=1$ for multiple thicknesses by means of the relative error $(w-w_{ref})/w_{ref}$ using one quadrilateral SBFEM or one/six SBFEM$_{ANS}$ elements for the discretization.}\label{tab: 4}%
\begin{tabular}{@{}llll@{}}
\toprule
 & \multicolumn{2}{@{}c@{}}{1 element} & \multicolumn{1}{@{}c@{}}{6 elements}\\\cmidrule(lr{0.5em}){2-3} \cmidrule(lr{0.5em}){4-4}
$t$ & SBFEM  & SBFEM$_{ANS}$ & SBFEM$_{ANS}$\\
\midrule
1& -58.4\%& 0\%& -5.76$\cdot10^{-15}$\% \\
        0.1& -99.3\%& 0\%& 3.59$\cdot10^{-12}$\% \\
        0.01&-99.99\%  & 4.36$\cdot10^{-9}$\% &-1.59$\cdot10^{-10}$\% \\
        0.001&-99.99\% &-3.23$\cdot10^{-8}$\% &7.83$\cdot10^{-9}$\%\\
\botrule
\end{tabular}
\end{table}
Next, the cantilever plate is discretized using six polygonal SBFEM$_{ANS}$ elements with three to seven nodes. The mesh is inspired by the polygonal patch test in \cite{Nguyen2017} and is displayed in Fig.\ref{fig: 6}. The coordinates of the nodes are given by
\begin{subequations}
\begin{align}
    &x = \Bigg[ 0,\ 2,\ 2,\ 0,\ \frac{1}{3},\ \frac{3}{2},\ \frac{4}{3},\ \frac{2}{3},\ \frac{3}{4},\ \frac{5}{4},\ 1\Bigg] ,\nonumber \\
        &y = \Bigg[0,\ 0,\ 1, \ 1, \ \frac{1}{6}, \ \frac{1}{4}, \ \frac{2}{3}, \ \frac{2}{3}, \ \frac{1}{8}, \ \frac{7}{24}, \ \frac{3}{4} \Bigg].
\end{align}
\end{subequations}
Again the numerical results for the displacements at the free edge are compared to the analytical solution from Eq.\eqref{eq: Ref. solution moment loading}, see Tab.\ref{tab: 4}. For all the thicknesses $t$, the numerical results correspond to the reference solution, displaying only slight numerical errors. As a result, an application of the derived formulation on polygonal elements is possible. 

\subsubsection{Uniformly distributed load}\label{sec: unifromly distributed load}
In the second load case, a uniformly distributed load is applied to the cantilever plate. The reference solution for the displacements at the free edge are calculated in accordance with Timoshenko`s beam theory, yielding
\begin{equation}
    w_{ref} = -\frac{qL^{4}}{8EI}-\frac{qL^{2}}{2GA_{s}},
\end{equation}
with $G = \frac{E}{2(1+\nu)}$ and $A_{s} = kA$. Here, $A= Bt$ is the cross-section and $I=Bt^3/12$ the moment of inertia of the cantilever plate. 

During the evaluation process, the geometry is discretized using quadrilateral SBFEM elements, incorporating either the ANS-approach, or a reduced (R) or selective-reduced (SR) integration technique. In the convergence analysis, mesh refinement is applied in length-direction, while in width-direction, one element per side is used.  

\begin{figure*}
\centering
\begin{minipage}{0.49\linewidth}
 \includegraphics[width=\linewidth]{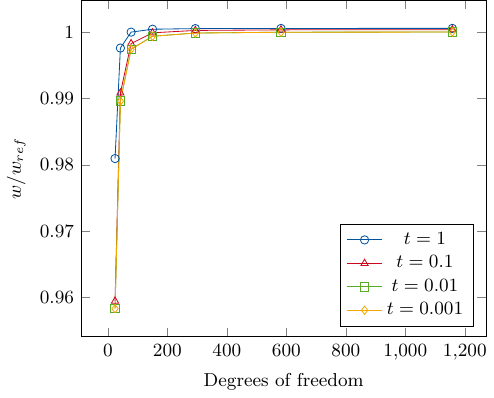}
    \centering (a)
\end{minipage}
\begin{minipage}{0.49\linewidth}
    \includegraphics[width=\linewidth]{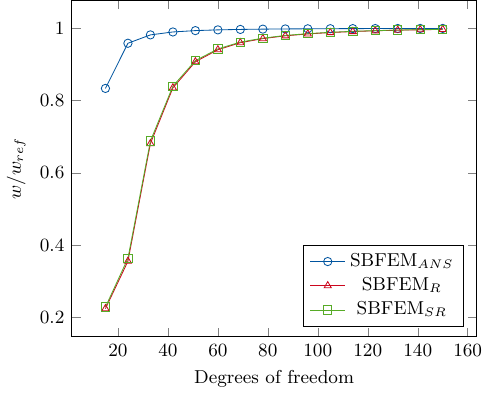}
    \centering (b)
\end{minipage}
\caption{(a) Convergence study of the relative displacement $w/w_{ref}$ at the free edge of a cantilever plate of thickness $t=\{1,\,0.1,\,0.01,\, 0.001\}$ discretized using quadrilateral SBFEM$_{ANS}$ elements and (b) comparison of the convergence of a cantilever plate of thickness $t=0.01$ discretized with SBFEM$_{ANS}$, SBFEM$_{R}$ and SBFEM$_{SR}$ elements, loaded by $q(x,y) = -10$.}
\label{fig: 7}
\end{figure*}

At first, the convergence behavior of the SBFEM$_{ANS}$ elements is analyzed using different thicknesses. The relative displacements $w/w_{ref}$ are plotted in dependence of the degrees of freedom in Fig.\ref{fig: 7}(a). It can be shown, that the displacements converge to the reference solution for all the chosen thickness values.

Using a plate of thickness $t=0.01$, the performance of the ANS formulation is compared to the behavior of SBFEM$_R$ and SBFEM$_{SR}$ in a convergence analysis. As previously, the relative displacement errors are evaluated at the free edge and displayed in dependence of the degrees of freedom in Fig.\ref{fig: 7}(b). The ANS converges faster than the reduced and selective reduced integration techniques.
\subsection{Clamped square plate}
\begin{figure*}
    \centering
    \input{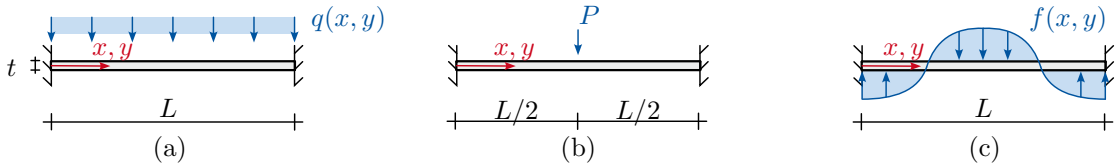}
    \caption{Illustration of the load cases (a) uniformly distributed load $q$, (b) central point load $P$ and (c) load function $f$ acting on the clamped square plate.}
    \label{fig: 8}
\end{figure*}

The next geometry is a clamped square plate with changing length $L$ and thickness $t$. In the following, three different load cases are examined. The first involves a uniformly distributed load $q(x,y)$. Then a point load $P$ is acting on the center of the plate. Last, a load function $f(x,y)$ is applied. The different load cases are visualized in Fig.\ref{fig: 8} for a cross-section of the plate. It follows a comparison of different mesh types, an analysis of Poisson's thickness locking in case of three-dimensional material laws, a test of the mesh sensitivity and an evaluation of the convergence rate.
\subsubsection{Uniformly distributed load} \label{sec: uniformly distributed load}
\begin{figure*}
    \centering
    \input{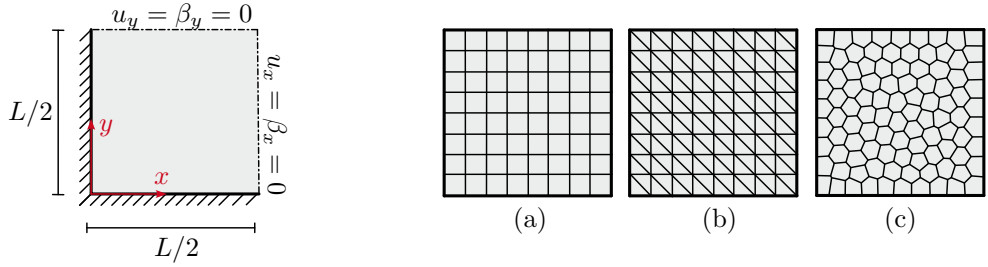}
    \caption{Boundary conditions due to symmetry of the clamped square plate and mesh types (a) quadrilateral, (b) triangular and (c) polygonal used for the discretization.}
    \label{fig: 9}
\end{figure*}

A fully clamped square plate of length $L=10$ and thickness $t=0.01$ is loaded by a uniformly distributed force $q(x,y)=-1$ in $z$-direction. In difference to the previous examples, the Poisson's ratio is non-zero and equal to $\nu=0.3$. The Young's modulus is set to $E= 10.92\cdot10^6$. As a result of the symmetry of the problem set-up, it is sufficient to discretize only one quarter of the plate. The boundary conditions due to symmetry are illustrated in Fig.\ref{fig: 9}.

During the evaluation process, the vertical displacement at the center of the plate ($x=L/2$, $y=L/2$) is compared to its analytical thin plate solution, see ~\cite{Weissmann1992,Timoshenko1959}. It can be determined as 
\begin{equation}
    w_{ref}= 0.00126\frac{qL^{4}}{D}=-12.6,
\end{equation}
with $D=Et^{3}/(12(1-\nu^{2}))$ being the flexural rigidity.

Considering the meshing, the quarter of the plate is discretized using triangular (T), quadrilateral (Q) and polygonal (P) SBFEM$_{ANS}$ elements, which allows their comparison. The polygonal meshes are generated using the $\mathtt{PolyMesher}$ by \cite{Talischi2012}. Examples of the meshes are visualized in Fig.\ref{fig: 9}.  

\begin{figure*}
\centering
\begin{minipage}{0.49\linewidth}
\input{Fig_10_a.tex}
    \centering (a)
\end{minipage}
\begin{minipage}{0.49\linewidth}
    \includegraphics[width=\linewidth]{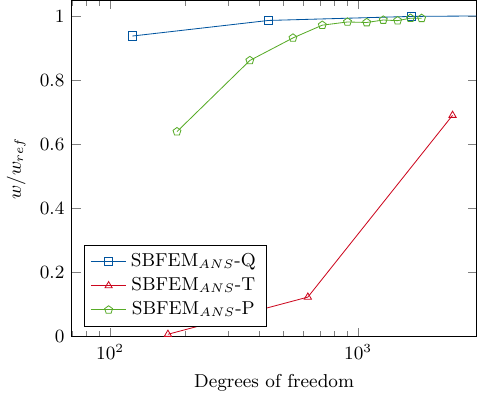}
   \centering (b)
\end{minipage}
\caption{(a) Illustration of the distribution of $w$ of a quarter of the clamped square plate under a uniformly distributed load $q(x,y) = -1$, using $16\times 16$ quadrilateral SBFEM$_{ANS}$ elements. (b) Convergence analysis of the relative displacement $w/w_{ref}$ at the center of the clamped square plate under the uniformly distributed load, discretized using triangular (T),  quadrilateral (Q) or polygonal (P) SBFEM$_{ANS}$ elements.}
\label{fig: 10}
\end{figure*}

The convergence behavior of the three mesh types is compared in Fig.\ref{fig: 10}(b). It can be observed, that the quadrilateral elements are the most suitable elements for the discretization of the square plate. However, the polygonal elements also converge quite fast, while the triangular elements behave much stiffer. In Fig.\ref{fig: 10}(a), the results for the displacement distribution $w$ of a mesh of $16\times 16$ quadrilateral elements are displayed.
\subsubsection{Mesh distortion test}
\begin{figure*}
    \centering
    \input{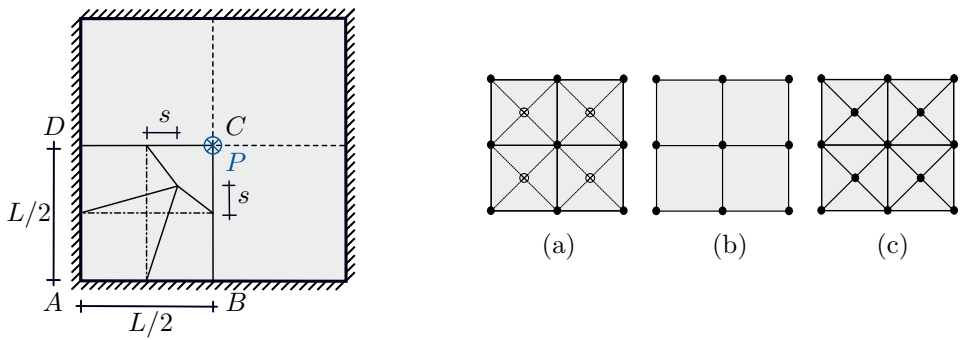}
    \caption{Problem set-up of the mesh distortion test and undistorted meshes of one quarter of a clamped square plate discretized using 4 SBFEM$_{ANS}$ (a), 4 Q1$_{ANS}$ (b) or 16 MT3 elements (c).}
    \label{fig: 11}
\end{figure*}

In numerical methods, sensitivity towards mesh distortion can compromise the performance of element formulations. It is therefore of importance to analyze the mesh sensitivity of the newly derived SBFEM$_{ANS}$ formulation. For this, a clamped square plate under a concentrated load $P=-16.367$ is typically used. As in \cite{Li2015}, the plate has a length of $L=100$ and a thickness $t=1$. Meanwhile, the Young's modulus is equal to $E=10^{4}$ and the Poisson's ratio is given by $\nu = 0.3$. Due to symmetry, it is sufficient to discretize one quarter of the plate, wherefore four quadrilateral elements are used, as displayed in Fig.\ref{fig: 11}. The boundary conditions are given by $w=\beta_{x}=\beta_{y}=0$ along edges AB and AD, $\beta_{x}=0$ along edge BC, and $\beta_{y}=0$ along edge DC. Initially the mesh is undistorted and consists of four uniform quadrilaterals. The distortion is incrementally increased by displacing the central node of the mesh by the parameter $s$. 
In the evaluation process, the numerical results for the vertical displacement are compared to a reference solution from literature \cite{Timoshenko1959}, defined as
\begin{equation}
    w_{ref} = 0.0056\frac{PL^{2}}{D} = -1.  \label{eq: wref distortion test}
\end{equation}
\begin{figure*}
\centering
\begin{minipage}{0.49\linewidth}
\input{Fig_12_a.tex}
    \centering (a)
\end{minipage}
\begin{minipage}{0.49\linewidth}
    \includegraphics[width=\linewidth]{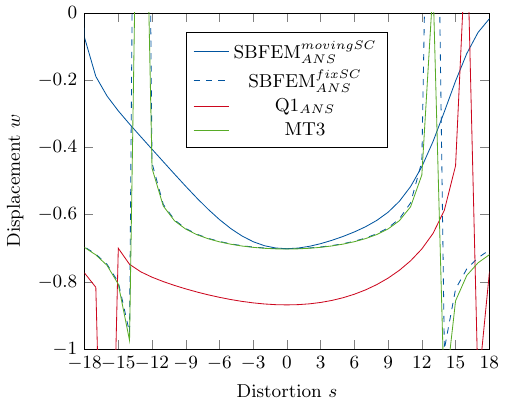}
    \centering (b)
\end{minipage}
\caption{(a) Illustration of the distorted SBFEM$_{ANS}$ elements with moving or fixed scaling centers (SC) and distortion $s=12.5$. (b) Vertical displacement $w$ at point $C$ of a clamped square plate under a point load $P$ in dependence of the distortion $s$ of the mesh.}
\label{fig: 12}
\end{figure*}

The mesh sensitivity of the SBFEM$_{ANS}$ formulation is compared to two other low-order plate formulations using assumed natural strains as a locking remedy. The first one is a quadrilateral bilinear element Q1$_{ANS}$ based on \cite{Bathe1985}. The second one is a triangular element MT3 \cite{Pal-Gap1999}. Further, a distinction is made in dependence of the scaling centers (SC) for the SBFEM elements. Their location is either fixed to $x_{0} = \big[ 12.5,\ 37.5,\ 12.5,\ 37.5\big]$ and $y_{0} = \big[ 12.5,\ 12.5,\ 37.5,\ 37.5\big]$ or variable. The location of the variable scaling centers is computed to the center of mass of each element. The undistorted meshes used for the evaluation are depicted in Fig.\ref{fig: 11}.

In Fig.\ref{fig: 12}(b), the vertical displacement $w$ under the point load is plotted in dependence of the mesh distortion $s$. It can be observed that in case of distortions of $|s|< 16$, the Q1$_{ANS}$ elements display the best performance, reaching a displacement of $w = -0.86$. However, the formulation fails in case of a distortion $s=16$. Meanwhile, SBFEM$_{ANS}^{\text{fix SC}}$ formulation displays nearly identical behavior to the MT3 formulation. Both fail in case of $12.5\leqslant|s|\leqslant 13$. Considering the SBFEM formulation, this is due to the scaling center coinciding with the displaced node and can be avoided by choosing variable scaling centers, as displayed in Fig.\ref{fig: 12}(a). The SBFEM$_{ANS}^{\text{moving SC}}$ show a slightly higher mesh sensitivity than the SBFEM$_{ANS}^{\text{fix SC}}$ elements. Though, they are applicable for a distortion of $12.5\leqslant|s|\leqslant 13$.
\subsubsection{Poisson's thickness locking analysis} \label{sec: Poisson's thickness locking analysis}
As mentioned in Section \ref{sec: Incorporation of 3D material laws}, the effect of Poisson's thickness locking occurs when considering three-dimensional material laws. To visualize the effect, a convergence analysis is performed on a clamped square plate using quadrilateral SBFEM$_{ANS}$ with $\mathbf{A}_{2}$ or $\mathbf{A}_{2}^{*}$ for the interpolation of the thickness strains, see Eq.\eqref{eq: A1 and A2}. The plate has a length of $L=100$ and a thickness $t=1$. The Young's modulus is equal to $E=10^4$ and the Poisson's ratio is first set to $\nu=0.3$ and then $\nu= 0$. As in the previous example, a concentrated load $F=-16.367$ is applied. The reference solutions for the central deflection at point $C$ are derived from Eq.\eqref{eq: wref distortion test}, yielding $w_{ref1} = -1$ for $\nu =0.3$ and $w_{ref2} = -1.0998624$ for $\nu=0$.

\begin{figure*}
\centering
\begin{minipage}{0.49\linewidth}
\includegraphics[width=\linewidth]{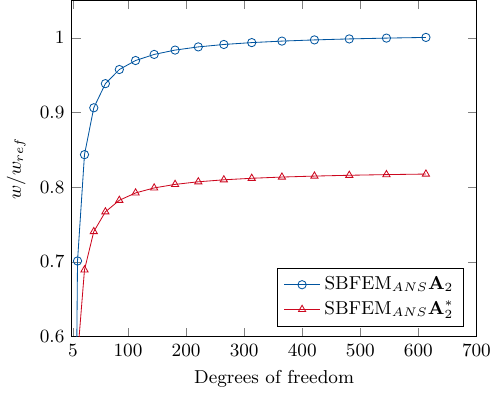}
    \centering(a)
\end{minipage}
\begin{minipage}{0.49\linewidth}
\includegraphics[width=\linewidth]{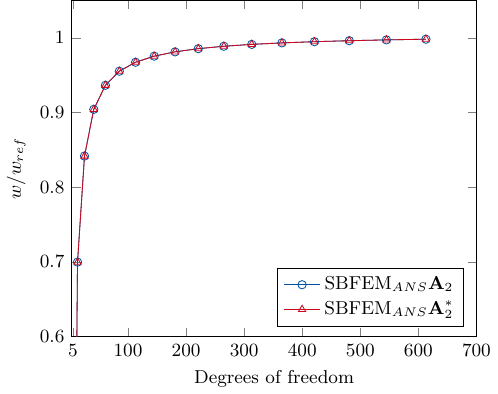}
    \centering (b)
\end{minipage}
\caption{Relative displacement $w/w_{ref}$ at point $C$ of the clamped square plate with a Poisson's ratio $\nu_{1} = 0.3$ (a) and   $\nu_{2} = 0$ (b) discretized using SBFEM$_{ANS}$ elements with a linear ($\mathbf{A}_{2}$) and a constant ($\mathbf{A}_{2}^{*}$) interpolation in thickness direction.}
\label{fig: 13}
\end{figure*}

In Fig.\ref{fig: 13}(a), the relative displacements $w/w_{ref}$ of a plate with $\nu=0.3$ are displayed in dependence of the degrees of freedom. It can be observed that, the formulation with a linear interpolation in thickness direction ($\mathbf{A}_{2}$) converges to the reference solution, while the formulation with a constant interpolation ($\mathbf{A}_{2}^{*}$) converges to $82\%$ of the reference solution. In the case of zero Poisson's ratio, both formulations behave identically, displaying good convergence behavior, as depicted in Fig.\ref{fig: 13}(b). This dependency on the value of $\nu$ is typical for Poisson's thickness locking and can be overcome by a linear interpolation of thickness strains in thickness direction, meaning the use of $\mathbf{A}_{2}$.

For all the examples, the results for SBFEM$_{ANS}\mathbf{A}_{2}$ and SBFEM$_{ANS}$ have proven to be identical. Therefore, the two dimensional material laws will be used for the following examples.

\subsubsection{Convergence rate evaluation} \label{sec: Convergence rate evaluation}
\begin{figure*}
\centering
\begin{minipage}{0.49\linewidth}
\input{Fig_14_a.tex}
    \centering (a)
\end{minipage}
\begin{minipage}{0.49\linewidth}
\includegraphics[width=\linewidth]{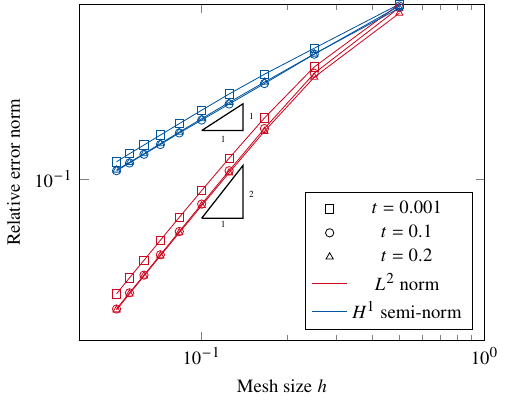}
    \centering(b)
\end{minipage}
\caption{(a) Illustration of the load function $f(x,y)$ for the convergence rate analysis of a clamped square plate. (b) Rates of convergence of the relative error norms of the clamped square plate loaded  by the load function $f(x,y)$.}
\label{fig: 14}
\end{figure*}
A clamped square plate of length $L=1$ and a variating thickness $t= \{ 0.2,0.1,0.001\}$ is loaded by a surface load defined by the following function
\begin{multline}
    f(x,y) = \frac{Et^{3}}{12(1-\nu^2)}\Big[ 12y(y-1)(5x^{2}-5x+1)\cdot\\\Big(2y^{2}(y-1)^{2}+x(x-1)(5y^{2}-5y+1)\Big) \\+ 12x(x-1)(5y^{2}-5y+1) \cdot\\\Big(2x^{2}(x-1)^{2} + y(y-1)(5x^{2}-5x+1)\Big)\Big].\label{eq: load function}
\end{multline}
This  function acts in thickness direction of the plate. A qualitative illustration of the function over the plate domain can be seen in Fig.\ref{fig: 14}(a). The material parameters are given by Poisson's ratio $\nu=0.3$ and $E = 1.092\cdot10^{7}$. An analytical solution for the displacements and the rotations can be found in \cite{Veiga2012} and is given by
\begin{multline}
    w_{ref} = \frac{1}{3}x^{3}(x-1)^{3}y^{3}(y-1)^{3}\\-\frac{2t^{2}}{5(1-\nu)} \cdot\Big[y^{3}(y-1)^{3}x(x-1)\cdot(5x^{2}-5x+1)\\+x^{3}(x-1)^{3}y(y-1)(5y^{2}-5y+1)\Big] \label{eq: Ref disp}
\end{multline} 
for the vertical displacements and 
\begin{equation}
    \boldsymbol{\beta}_{ref} = \begin{bmatrix}
    -(y^{3}(y-1)^{3}x^{2}(x-1)^{2}(2x-1))\\
        x^{3}(x-1)^{3}y^{2}(y-1)^{2}(2y-1) \label{eq: Ref rotations}
    \end{bmatrix}
\end{equation}
for the rotations around the $x$- and $y$-axis. 
To evaluate the SBFEM$_{ANS}$ formulation, the relative $L^{2}$ error norm and the relative $H^{1}_{s}$ semi-norm of the error are calculated. These norms are derived using the following formulas
\begin{subequations}
    \begin{align}
    &\frac{||\mathbf{v}-\mathbf{v}^{h}||_{L^{2}(\Omega)}}{||\mathbf{v}||_{L^{2}(\Omega)}} = \sqrt{\frac{\int_{\Omega}(\mathbf{v}-\mathbf{v}^{h})\cdot (\mathbf{v}-\mathbf{v}^{h})\,\text{d}\Omega} {\int_{\Omega}\mathbf{v}\cdot \mathbf{v}\,\text{d}\Omega}}, \\ & \frac{||\mathbf{v}-\mathbf{v}^{h}||_{H^{1}_{s}(\Omega)}}{||\mathbf{v}||_{H^{1}_{s}(\Omega)}} = \sqrt{\frac{\int_{\Omega}(\mathbf{v}'-\mathbf{v}'^{h})\cdot (\mathbf{v}'-\mathbf{v}'^{h})\,\text{d}\Omega} {\int_{\Omega}\mathbf{v}'\cdot \mathbf{v}'\,\text{d}\Omega}}, \label{eq: error norms}
\end{align}
\end{subequations}
where $\mathbf{v}$ is defined as $\mathbf{v} = \begin{bmatrix}
    w & \beta_{x}& \beta_{y}
\end{bmatrix}^{\text{T}}$ and $\mathbf{v}'$ contains the partial derivatives with respect to $x$ and $y$ such that $\mathbf{v}' = \begin{bmatrix}
     w_{,x} &w_{,y} & \beta_{x,x}&  \beta_{x,y}&\beta_{y,x} & \beta_{y,y}
\end{bmatrix}^{\text{T}}$.

For the numerical analysis, the square plate is discretized using uniform meshes of quadrilateral SBFEM$_{ANS}$ elements, where the number of elements per side variates from 2 to 20 during the evaluation. In Fig.\ref{fig: 14}(b), the relative error norms are displayed in dependence of the mesh size $h$, which is defined as $h=1/\sqrt{n_{elem}}$, with $n_{elem}$ being the total amount of elements in the mesh. 
Independent of the slenderness of the plate, the optimal convergence rate is obtained for both error norms. This means that transverse shear locking has successfully been overcome using the ANS approach.
\subsection{Simply supported square plate}
\begin{figure}
    \centering
    \input{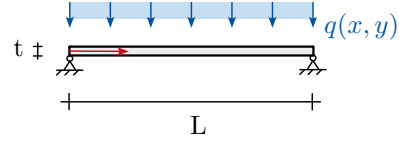}
    \caption{Illustration of the investigated load case of the simply supported square plate.}
    \label{fig: 15}
\end{figure}

\begin{figure}
    \centering
    \input{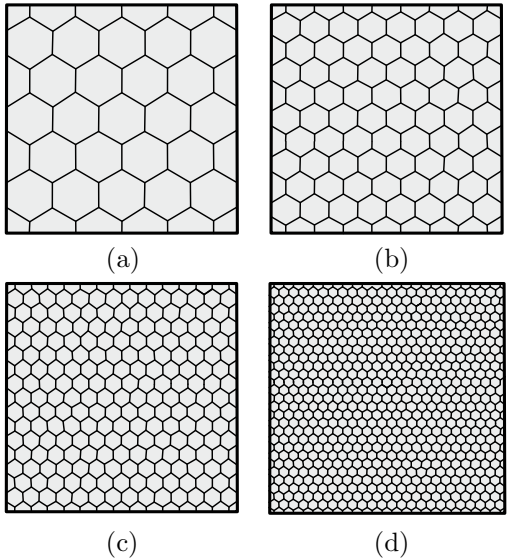}
    \caption{Mesh of (a) 39 , (b) 94 ,(c) 247 and (d) 711 polygonal elements used for the discretization of the quarter of the simply supported square plate.}
    \label{fig: 16}
\end{figure}

\begin{figure*}
\centering
\begin{minipage}{0.49\linewidth}
 \includegraphics[width=\linewidth]{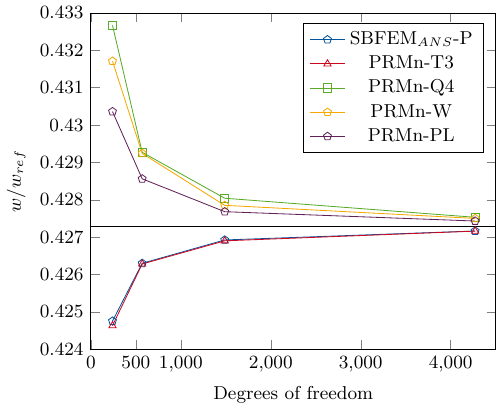}
    \centering (a)
\end{minipage}
\begin{minipage}{0.49\linewidth}
    \includegraphics[width=\linewidth]{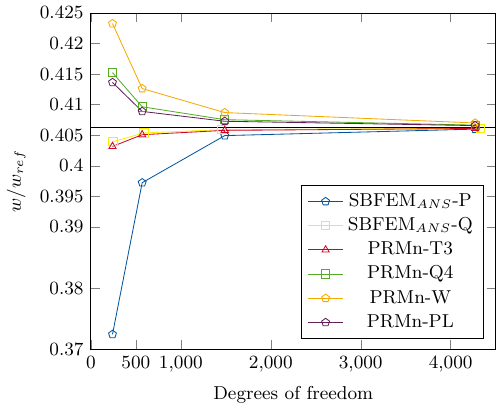}
    \centering (b)
\end{minipage}
\caption{Convergence study of the central deflection $\delta$ of a simply supported square plate of thickness (a) $t=1$ and (b) $t = 0.01$ discretized using quadrilateral or polygonal SBFEM$_{ANS}$ elements and comparison to the results of triangular, quadrilateral and polygonal elements with barycentric shape functions from \cite{Nguyen2017}.}
\label{fig: 17}
\end{figure*}

A simply supported square plate is loaded by a uniformly distributed force $q(x,y) = 1$ as displayed in Fig.\ref{fig: 15}. Due to the symmetry of the problem set-up, it is sufficient to discretize one quarter of the plate of a length of $L/2=10$ and a thickness equal to $t=\{1,\,0.01\}$. As in \cite{Nguyen2017}, the material parameters are given by $E = 10.92\cdot 10^5$ and $\nu = 0.3$. During the evaluation process, the results for the central deflection defined as:
\begin{equation}
    \delta = \frac{100Dw_c}{qL^4}
\end{equation}
are compared to results from literature ~\cite{Taylor1993,Zienkiewicz1993}. The reference value for the deflection of a plate with $t = 1$ is given by $0.42728$ and for a plate of thickness $t=0.01$ it is equal to $0.40624$.  Further a comparison of the convergence behavior of polygonal (-P) and quadrilateral (-Q) SBFEM$_{ANS}$ elements to other polygonal Reissner--Mindlin plate elements using two different barycentric shape function types is carried out. These reference elements from \cite{Nguyen2017} use Wachspress (PRMn-W) and piecewise-linear (PRMn-PL) functions. Their derivation and numerical results of the central deflection for the simply supported square plate using four different meshes are given in \cite{Nguyen2017}. The polygonal meshes considered are displayed in Fig.\ref{fig: 16} and consist of 39, 94, 247 or 711 elements. Additionally, a triangular (PRMn-T3) and a quadrilateral (PRMn-Q4) element formulation from \cite{Nguyen2017} are considered, where 8, 13, 21 and 37 elements per side are used. The results for the plate of thickness $t = 1$ are displayed in Fig.\ref{fig: 17}(a). It can be observed that both the PRMn-T3 and the SBFEM$_{ANS}$-P elements converge from below towards the reference value, while the other formulations converge from above. As shown, the SBFEM$_{ANS}$-P elements behave almost identically to the PRMn-T3 elements. In Fig.\ref{fig: 17}, the results for the slender plate of thickness $t=0.01$ are given. Here, a discrepancy in dependence of the meshes and element shapes could be deduced for the SBFEM$_{ANS}$ elements, similar to Section \ref{sec: uniformly distributed load}.
\subsection{Clamped circular plate}
\begin{figure*}
    \centering
    \input{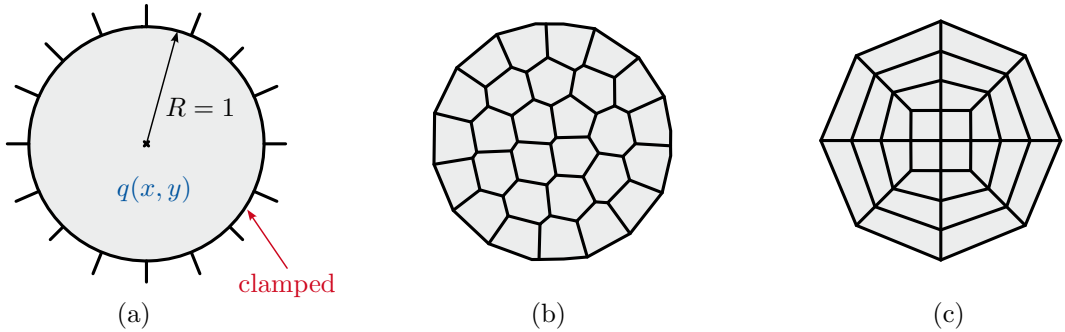}
    \caption[Geometry of the clamped square plate.]{(a) Geometry and boundary conditions of the clamped circular plate and domain discretized with \cite{Talischi2012} using 28 polygonal elements (b) or quadrilateral elements (c).}
    \label{fig: 18}
\end{figure*}

A circular plate of radius $R=1$ is clamped at the outer boundary as illustrated in Fig.\ref{fig: 18}(a). It is loaded by a uniformly distributed force $q = 1$ in $z$-direction. The Young's modulus is $E = 10.92\cdot10^6$ and the Poisson's ratio is $\nu = 0.3$. An analytical derivation of the solutions for the displacements and rotations can be found in ~\cite{Videla2019,Katili1993}. They are given by
\begin{multline}
    w_{ref} = \frac{(x^2+y^2)^2}{64D}-(x^2+y^2)\cdot\Bigg[\frac{t^2}{4\lambda}+\frac{1}{32D}\Bigg]\\+\frac{t^2}{4\lambda} + \frac{1}{64D},
\end{multline}
\begin{equation}
    \boldsymbol{\beta}_{ref} =\frac{x^2+y^2-1}{16D} \begin{bmatrix}
    -x\\
      y
    \end{bmatrix},
\end{equation}
where $t$ is the thickness chosen to $t=\{0.1,0.2\}$ during the evaluation process. As in \cite{Videla2019}, the parameter $\lambda$ is defined as $\lambda = \frac{\kappa E t^3}{2(1+\nu)}$ and $D$ is the flexural rigidity.

The circular plate is discretized by polygonal SBFEM$_{ANS}$ elements, using $\mathtt{PolyMesher}$ by \cite{Talischi2012}. Meshes with 16, 32, 64, ..., 1024 elements are considered. An example mesh with 28 polygonal elements is displayed in Fig.\ref{fig: 18}(b). For each of these meshes the relative $L^{2}$ error norm and the $H^{1}_{s}$ semi-norm of errors are evaluated, using Eq.\eqref{eq: error norms}.

\begin{figure*}
\centering
\begin{minipage}{0.49\linewidth}
\includegraphics[width=\linewidth]{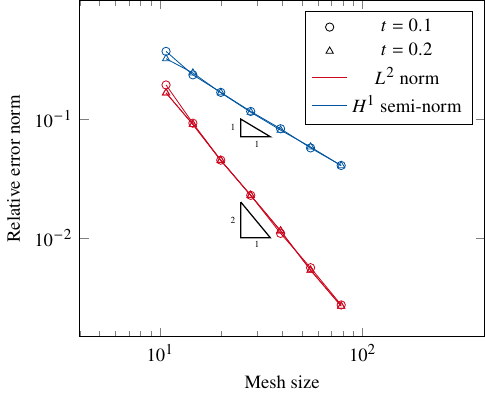}
    \centering (a)
\end{minipage}
\begin{minipage}{0.49\linewidth}
\includegraphics[width=\linewidth]{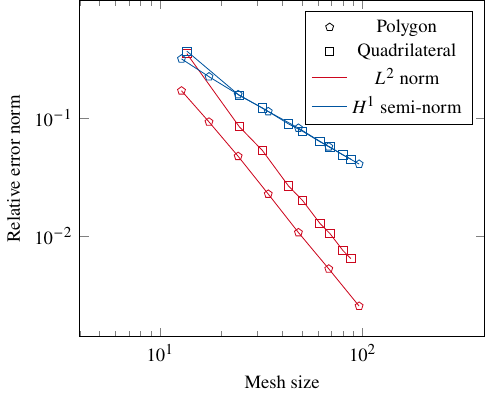}
    \centering (b)
\end{minipage}
\caption{(a) Rates of convergence of the relative error norms for the clamped circular plate of thickness $t=\{0.1,0.2\}$ discretized using polygonal SBFEM$_{ANS}$ elements. (b) Rates of convergence of the relative error norms for the clamped circular plate of thickness $t=0.1$ discretized using polygonal and quadrilateral SBFEM$_{ANS}$ elements.}
\label{fig: 19}
\end{figure*}

In Fig.\ref{fig: 19} (a), the relative error norms are displayed in dependence of the mesh size for the two thicknesses $t=0.1$ and $t=0.2$.  The expected optimal convergence rates are reached for both norms, independent of the slenderness of the plate. This proves a successful alleviation of transverse shear locking.

Further, the clamped circular plate is discretized using quadrilateral elements. This allows the comparison of the mesh types on a more complex geometry. An example mesh of the quadrilateral discretization is given in Fig.\ref{fig: 18}(c). The results for the relative $L^2$ and $H^1_s$ error norms of a plate of thickness $t=0.1$ are displayed in Fig.\ref{fig: 19}(b). It can be observed that the polygonal discretization is slightly better than the quadrilateral discretization in case of the $L^2$ norm.

\subsection{L-Bracket}
\begin{figure*}
    \centering
    \input{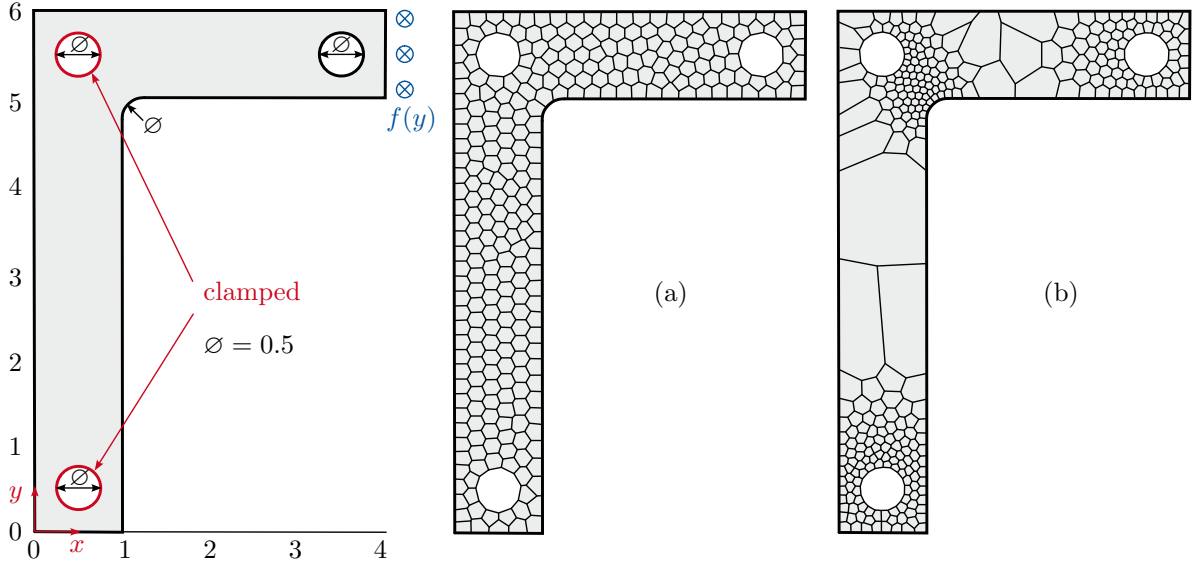}
    \caption[Geometry of the clamped square plate.]{Geometry and boundary conditions of the L-bracket and domain discretized by a uniform (a) and a locally refined (b) mesh of 300 polygonal elements.}
    \label{fig: 20}
\end{figure*}

To display the applicability of the formulation to complex geometries using a uniform or a locally refined mesh of polygonal elements, an L-Bracket with three circular holes is analyzed. Previous studies of this example appear in ~\cite{Arf2023,Coradello2021}. The bracket has a thickness $t=0.01$ and the geometric description can be found in Fig.\ref{fig: 20}. Neumann and Dirichlet boundary conditions are also visualized. The line load is constant and acting in vertical direction. It is given by $f(x=4,y)=-100$. The Young's modulus is $E= 200\cdot 10^9$ and the Poisson's ratio is $\nu = 0$. 

\begin{figure*}
\centering
\begin{minipage}{0.3\linewidth}
\input{Fig_21_a.tex}
\end{minipage}
\begin{minipage}{0.49\linewidth}
\includegraphics[width=\linewidth]{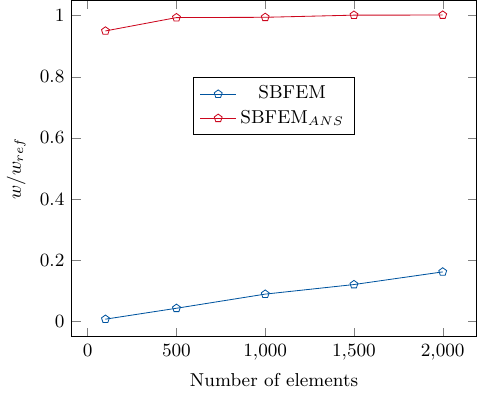}
    \centering (b)
\end{minipage}
\caption{(a) Numerical results of the distribution of the displacement $w$. (b) Convergence of the relative displacement $w/w_{ref}$ at point ($x=4$, $y=6$) of the L-Bracket discretized using a uniform mesh of polygonal SBFEM or SBFEM$_{ANS}$ elements in dependence of the number of elements.}
\label{fig: 21}
\end{figure*}

To evaluate the problem, the out-of-plane displacement $w$ at the node with the coordinates $x=4$ and $y=6$ is compared to the reference solution $w_{ref}=-0.07$ of the isogeometric Kirchhoff plate formulation in ~\cite{Arf2023}. At first, the performance of uniform meshes generated with $\mathtt{PolyMesher}$, as in Fig.\ref{fig: 20}(a), is analyzed. SBFEM$_{ANS}$ elements are compared to that of the SBFEM formulation without any locking remedy in Fig.\ref{fig: 21}(b). As shown, the SBFEM$_{ANS}$ performs significantly better and reaches the 
reference solution for the displacement with only 500 elements while the other formulation is still under $20\%$ of the reference solution with 2000 elements. 

\begin{figure*}
    \centering
    \input{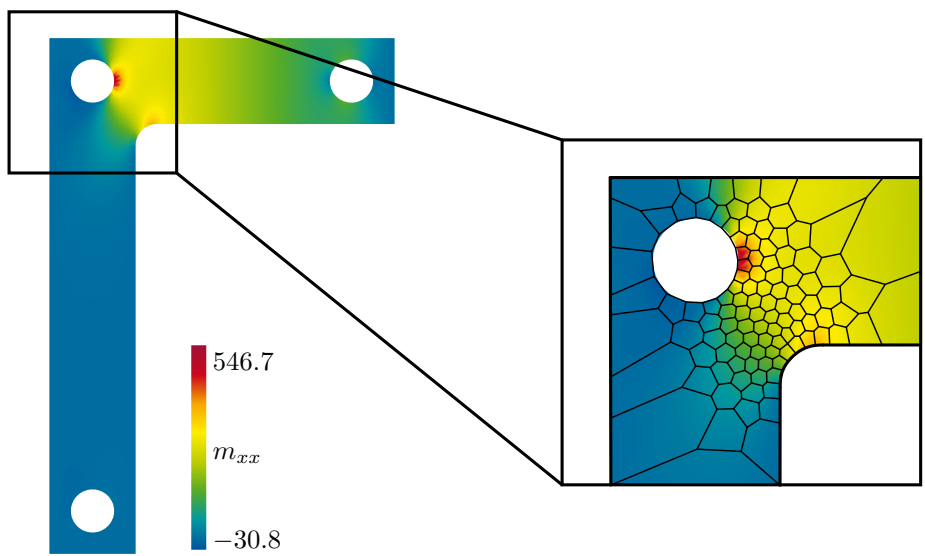}
    \caption{Numerical results for $m_{xx}$ using a uniform mesh of 3000 SBFEM$_{ANS}$ elements and proposed local refined mesh based on the stress distribution.}
    \label{fig: 22}
\end{figure*}

\begin{figure*}
    \centering
    \input{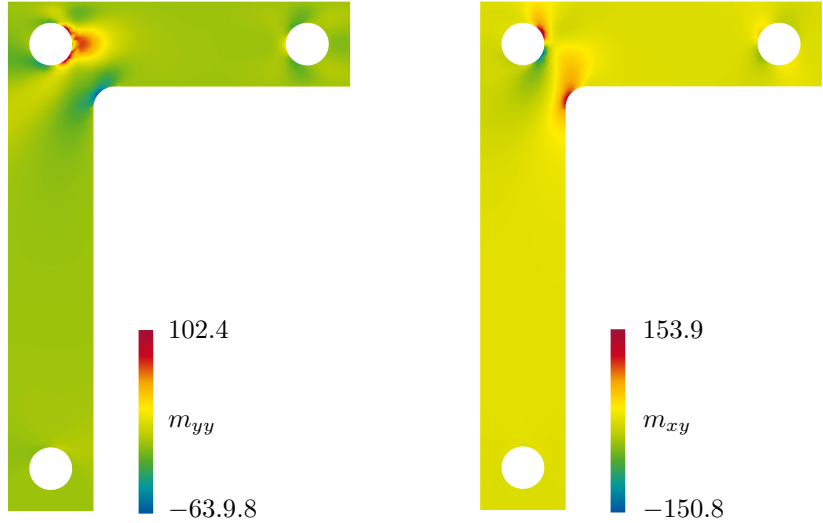}
    \caption{Numerical results for $m_{yy}$ and $m_{xy}$ using a uniform mesh of 3000 SBFEM$_{ANS}$ elements.}
    \label{fig: 23}
\end{figure*}

Next, the maximum stress $m_{xx}$ at $x=0.75$, $y=5.5$ is analyzed. Here, the reference value from Abaqus is given by $m_{xx,ref} = 599.64$, using a mesh of 161228 S8R STRI65 elements. This value agrees with the reference solution of $m_{xx,ref} =600$ reported in \cite{Coradello2021}. For a uniform mesh of 3000 SBFEM$_{ANS}$ elements, the distribution of $m_{xx}$ over the element domain is displayed in Fig.\ref{fig: 22} and those of $m_{xy}$ and $m_{yy}$ in Fig.\ref{fig: 23}. At $x=0.75$, $y=5.5$, a value of $m_{xx} = 489.1184$ is reached, which does not correspond to the reference solution yet. Meanwhile, the reference solution for the displacement is already reached for a much coarser mesh. To improve the results for the stresses, a locally refined mesh is introduced. 
At the holes and the point of interest for the stresses, the refinement is induced by a exponential density function defined as 
\begin{equation}
    d = exp\big(-\frac{1}{2\sigma^2}\big[(x-x_c)^2+(y-y_c)^2\big]\big), 
\end{equation}
with $0 < \sigma \leqslant 1.$

The coordinates ($x_c$, $y_c$) are those of the attractor points and $\sigma$ defines the ratio of the refinement. An example of a mesh of 300 elements can be seen in Fig.\ref{fig: 20} (b).

\begin{figure*}
\centering
\begin{minipage}{0.49\linewidth}
\includegraphics[width=\linewidth]{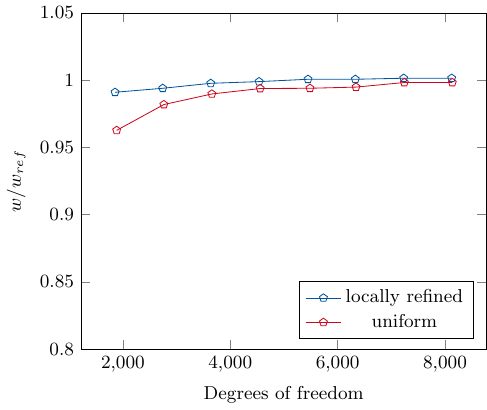}
    \centering (a)
\end{minipage}
\begin{minipage}{0.49\linewidth}
\includegraphics[width=\linewidth]{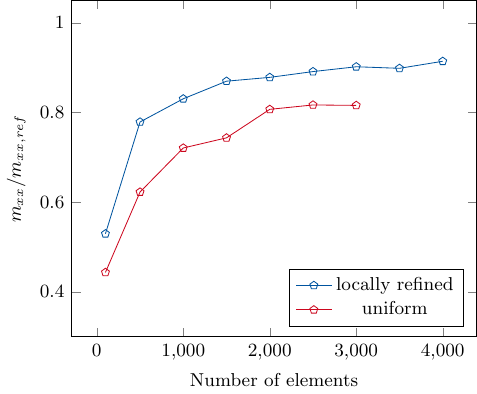}
    \centering (b)
\end{minipage}
\caption{(a) Convergence of the relative displacement $w/w_{ref}$ at point ($x=4$, $y=6$) of the L-Bracket discretized using a uniform or locally refined mesh of polygonal SBFEM$_{ANS}$ elements in dependence of the degrees of freedom. (b) Convergence of the relative stress $m_{xx}/m_{xx,ref}$ at point ($x=0.75$, $y=5.5$) of the L-Bracket discretized using a uniform or locally refined mesh of polygonal SBFEM$_{ANS}$ elements in dependence of the number of elements.}
\label{fig: 24}
\end{figure*}

The performance of uniform meshes is compared to those of locally refined meshes of SBFEM$_{ANS}$ elements. The results for the relative displacements $w/w_{ref}$ are displayed in dependence of the number of degrees of freedom in Fig.\ref{fig: 24}. The locally refined meshes show a faster convergence. However the discrepancies for the displacements are quite small. In case of the stress $m_{xx}$ at point $x=0.75$, $y=5.5$, the advantage of the locally refined mesh compared to the uniform mesh is more obvious. The refined meshes outperform the uniform ones with differences of $10\%$. Meanwhile, the reference value for the stresses is still not reached using up to 3000 polygonal elements and a mesh dependency could be deduced in the convergence of the stresses. Here, an error estimation would be of interest to optimize the mesh refinement further.  

\section{Conclusion and outlook}\label{sec: Conclusion and outlook}
In the face of the challenge posed by  discretizing increasingly complex plate geometries, this paper presents a polygonal Reissner--Mindlin plate formulation, using the framework of fully-discretized scaled boundary finite element methods. The common effect of transverse shear locking in the thin plate limit is addressed by an assumed natural strain approach, which was specifically derived for application on scaled boundary sections. Its effectiveness was tested in various benchmarking examples, with progressively more complex geometries and load cases. Here, the convergence to reference solutions from literature is significantly improved by the incorporation of assumed natural strains, displaying a successful alleviation of transverse shear locking. Investigations regarding the element shapes were further made, showing that, depending on the geometry, the use of quadrilateral or polygonal elements is more beneficial. Further, local mesh refinement techniques were used to discretize the L-bracket, where a reduction of the computational cost could be deduced, compared to the uniform meshes. However, no error estimator was incorporated yet and the improvements of the results through local refinement are slight. This underlines the importance of error estimators to reasonably refine meshes in the crucial regions, which may be of interest for future works. 

Regarding the material laws, the initial $2D$-continuum was extended to $3D$, through the incorporation of in-plane displacements and the introduction of thickness strains. Here, the necessity of a linear interpolation of the thickness strains in thickness direction to avoid Poisson's locking was highlighted in a numerical example. So far, the analysis was limited to linear, isotropic material models. However, the inclusion of more complex material models defined in $3D$-continuum is imaginable.

In general, it can be concluded that a fully-discretized SBFEM formulation with assumed natural strains has successfully been derived.

\section*{Declarations}
\begin{itemize}
\item Funding: The authors would like to thank the German Research Foundation (Deutsche Forschungsgemeinschaft – DFG) for financial support of the research unit FOR5492 “Polytope Mesh Generation and Finite Element Analysis Methods for Problems in Solid Mechanics” under project number 495926269.
\end{itemize}
\begin{appendices}

\section{Convergence rate evaluation of the stresses}\label{secA1}
In the previous numerical examples, it was primarily focused on the analysis of the displacements and rotations. In this appendix, the stresses are evaluated and discussed using the example from Section \ref{sec: Convergence rate evaluation}. The clamped square plate of thickness $t=0.2$ is loaded by the load function defined in Eq.\eqref{eq: load function}. Analytical solutions for the displacements and rotations are given in Eq.\eqref{eq: Ref disp} and Eq.\eqref{eq: Ref rotations} and allow the derivation of the analytical results for the strains and the stresses. Therefore, Eq.\eqref{Eq: Curvature Strains} and Eq.\eqref{Eq: Shear strain}, as well as Eq.\eqref{eq: Stresses} are used. It follows an evaluation of the numerical results for the strains and stresses by means of the energy ($s$-)norm given by:
\begin{equation}
    e_s = \frac{||s-s^{h}||_{L^{2}(\Omega)}}{||s||_{L^{2}(\Omega)}} = \sqrt{\frac{\int_{\Omega}(\mathbf{\epsilon}-\mathbf{\epsilon}^{h})\cdot (\mathbf{\sigma}-\mathbf{\sigma}^{h})\,\text{d}\Omega} {\int_{\Omega}\mathbf{\epsilon}\cdot \mathbf{\sigma}\,\text{d}\Omega}}.\label{eq: L2 error}
\end{equation}
The plate is discretized using 2 to 10 quadrilateral SBFEM$_{ANS}$ elements per side. The results can be seen in Tab.\ref{tab: A.4}. It can be observed that the results for the $e_s$-energy norm reduce slowly and in difference to the $L^2$ error norm and the $H^1$ semi-norm of the displacement error, the optimal convergence rate is not reached. 
\begin{table}[htbp]
\caption{Results for the energy norm of the clamped square plate loaded by $f(x,y)$.}\label{tab: A.4}%
\begin{tabular}{@{}llllll@{}}
\toprule
  Nelem\footnotemark[1] & 2 & 4& 6 & 8 & 10\\
\midrule
 $e_s$& 0.836&0.809& 0.750& 0.728&0.718\\
\botrule
\end{tabular}
\footnotetext[1]{Number of elements per side.}
\end{table}
It can be concluded that in case of the clamped square plate, despite significantly improving the results for the displacements, issues regarding the stresses remain and need further investigations.
\label{app1}

\end{appendices}
\bibliographystyle{unsrt}
\bibliography{sn-bibliography}

\end{document}